\documentclass[11pt,a4paper]{article}
\usepackage{jcappub}
\pdfoutput=1                     % when submitting pdflatex (i.e. with pdf, png or jpg images)

\usepackage[T1]{fontenc}    
\usepackage{graphicx}
\usepackage{mathdots}
\usepackage{bm}               % Bolded math symbols (e.g. bolded greek alphabet).
\usepackage{slashed}

\usepackage{mathtools}        % Fixes and extends (and also loads) amsmath.
\usepackage{bbm}              % Blackboard math variants of CM fonts. (For \mathbbm{1}.)
\usepackage{bm}               % Bolded math symbols (e.g. bolded greek alphabet).
\usepackage{amsmath, amssymb} 
\usepackage{bbold}
\usepackage{tikz} % graphics macro package (load after hyperref, apparently)
\usepackage{feynmp-auto} % feynman diagrams w/ automatic metapost-processing
\usetikzlibrary{positioning,arrows.meta,decorations.markings,decorations.pathmorphing}
\usepackage{upgreek} % for upright pi \uppi
\usepackage[titletoc]{appendix}

\usepackage{cleveref} % better cross-refs using \cref (load after hyperref!)
    \crefname{equation}{}{}
    \crefname{figure}{}{}    % example: \crefformat{figure}{#figure~#1#3}
    \crefname{table}{}{}
    \crefname{section}{}{}   % example: \crefformat{section}{#2section~#1#3}
    \crefname{appendix}{}{}
    \crefname{footnote}{}{}
    
    \crefalias{subequation}{equation}

\def\nc{\newcommand}
\def\FD{{\scriptscriptstyle {\rm FD}}}
\def\MB{{\scriptscriptstyle {\rm MB}}}

\def\PA{{\scriptscriptstyle ||}}

\def\tsum{{\textstyle \sum}}

\nc{\be}{\begin{equation}}
\nc{\ee}{\end{equation}}
\nc{\bea}{\begin{eqnarray}}
\nc{\eea}{\end{eqnarray}}
\nc{\ba}{\begin{array}}
\nc{\ea}{\end{array}}

\nc{\nn}{\nonumber}

\nc{\deldag}{{\mathbin{\partial\mkern-10mu/}}}
\nc{\kdag}{{\mathbin{k\mkern-10mu\big/}}}
\nc{\Ddag}{{\mathbin{D\mkern-10mu\big/}}}

\def\m2p{{|m|^{2\prime}}}
\def\x2p{{|x|^{2\prime}}}

\def\Slashnew#1{#1\kern-0.55em\raise.05ex\hbox{/}}
\def\slashnew#1{#1\kern-0.5em\raise.05ex\hbox{{$\scriptstyle /$}}}

\def\lsim{\mathrel{\raise.3ex\hbox{$<$\kern-.75em\lower1ex\hbox{$\sim$}}}}
\def\gsim{\mathrel{\raise.3ex\hbox{$>$\kern-.75em\lower1ex\hbox{$\sim$}}}}

\nc{\shalf}{\ensuremath{\textstyle \frac{1}{2}}}
\nc{\ihalf}{\ensuremath{\textstyle \frac{i}{2}}}

\def\sfrac#1#2{\ensuremath{\textstyle \frac{#1}{#2}}}

\def\emph#1{{\em #1}}

\def\eg{{\em e.g.}}

\nc{\sss}{\scriptscriptstyle}
\nc{\W}{{\sss W}}
\nc{\sparallel}{{\sss\parallel}}
\nc{\tGamma}{{\tilde \Gamma}}

\def\R{{\rm R}}
\def\L{{\rm L}}
\def\tL{{t_{\rm -}}}
\def\bL{{b_{\rm -}}}

\def\tR{{t_{\rm +}}}

\nc{\HF}{{\sss \rm FH}}

\def\MB{{\scriptscriptstyle \rm MB}}

\nc{\flux}{\text{flux}}
\nc{\mol}{mol}
\nc{\intp}[1]{\int \frac{d^3 p_{#1}}{( 2\pi)^3 2 E_{#1}}}
\nc{\Mel}{\mathcal{M}}

\title{Systematic moment expansion for electroweak baryogenesis}

\author[a,b]{Kimmo Kainulainen}
\author[a]{Niyati Venkatesan}

\affiliation[a]{Department of Physics, PL 35 (YFL), 40014 University of Jyv\"askyl\"a, Finland}
\affiliation[b]{Helsinki Institute of Physics, PL 64, 00014 University of Helsinki, Finland}

\emailAdd{kimmo.kainulainen@jyu.fi}
\emailAdd{niyati.a.venkatesan@jyu.fi}

\abstract{We present a systematic moment expansion for solving the semiclassical Boltzmann equations for electroweak baryogenesis. The expansion is developed in powers of adiabatic coordinate velocity, and it is used for computing the CP-violating seed asymmetry at the front of the phase transition wall, that sources the eventual baryon asymmetry of the universe (BAU). We implement the method in a benchmark model, with a CP-violating mass arising from a dimension-5 operator coupling the fermion to a singlet scalar field. We find  that the higher moment calculations yield a BAU that can significantly differ from the commonly used two-moment approximation. We discuss in detail the underlying approximations in the moment method and propose a new truncation scheme for the expansion, that appears to give more numerically robust results than the previous schemes.}


\arxivnumber{}

\begin{document}
\maketitle

%%%%%%%%%%%%%%%%%%%%%%%%%%%%%%%%%%%%%%%%%%%%%%%%%%%%%%%%%%%%%%%%%%%%%%%%%%%%%%%%%%%%%%%%%%%%%%%%%%%
%
% --------------------- Article ----------------------------
%
%%%%%%%%%%%%%%%%%%%%%%%%%%%%%%%%%%%%%%%%%%%%%%%%%%%%%%%%%%%%%%%%%%%%%%%%%%%%%%%%%%%%%%%%%%%%%%%%%%%

%%%%%%%%%%%%%%%%%%%%%%%%%%%%%%%%%%%%%%%%%%%%%%%%%%%%%%%%%%%%%%%%%%%%%%%%%%%%%%%%%%%%%%%%%%%%%%%%%%%
%%%%%%%%%%%%%%%%%%%%%%%%%%%%%%%%%%%%%%%%%%%%%%%%%%%%%%%%%%%%%%%%%%%%%%%%%%%%%%%%%%%%%%%%%%%%%%%%%%%
%
\section{Introduction}
\label{sec:intro}
%
%%%%%%%%%%%%%%%%%%%%%%%%%%%%%%%%%%%%%%%%%%%%%%%%%%%%%%%%%%%%%%%%%%%%%%%%%%%%%%%%%%%%%%%%%%%%%%%%%%%
%%%%%%%%%%%%%%%%%%%%%%%%%%%%%%%%%%%%%%%%%%%%%%%%%%%%%%%%%%%%%%%%%%%%%%%%%%%%%%%%%%%%%%%%%%%%%%%%%%%

The origin of the baryon asymmetry in the Universe is still unknown. One possible explanation is
the electroweak baryogenesis (EWBG) mechanism, which relies on the nonperturbative baryon number violation in the standard model (SM)~\cite{Kuzmin:1985mm,Arnold:1987mh}. This mechanism requires that the electroweak phase transition (EWPT) is of the first order~\cite{Bochkarev:1990fx,Cohen:1990py,Cohen:1990it,Turok:1990zg}, while in the SM the transition is a smooth crossover~\cite{Kajantie:1996mn}. A first order EWPT can, however, be realized in many beyond standard model (BSM) scenarios, and a strong first order phase transition would also be an interesting source of gravitational waves~\cite{Vaskonen:2016yiu,Bian:2017wfv,Dorsch:2016nrg,Huang:2016cjm,Artymowski:2016tme,Hashino:2016xoj,Chao:2017vrq,Beniwal:2017eik,Kurup:2017dzf,Angelescu:2018dkk,Beniwal:2018hyi,Ahriche:2018rao,Huang:2018aja}.

A central problem in the EWBG mechanism is the computation of the CP-violating seed asymmetry around expanding phase transition fronts, which biases the B-violating interactions to create the baryon asymmetry in the universe (BAU). Historically, two distinct approaches existed to this problem: the VIA-method (see \eg~\cite{Postma:2019scv}) and the semiclassical force mechanism (SC)~\cite{Joyce:1994zt,Cline:2000nw,Kainulainen:2001cn,Kainulainen:2002th,Prokopec:2003pj,Prokopec:2004ic,Fromme:2006wx,Cline:2017qpe,Cline:2020jre}, but it was recently shown that the VIA-results are incorrect~\cite{Kainulainen:2021oqs,Postma:2022dbr}. In this paper we use the semiclassical (SC) method, where the CP-violation emerges via a semiclassical force in the Boltzmann equations for particle distribution functions. The SC equations have been derived using WKB-methods~\cite{Cline:1997vk,Cline:2000nw,Cline:2001rk,Kainulainen:2001cn} and from the closed-time-path (CTP) formalism~\cite{Kainulainen:2001cn,Kainulainen:2002th,Prokopec:2003pj,Prokopec:2004ic}. For the latest treatment, with an extension to include thermal corrections to the SC-force, see~\cite{Kainulainen:2021oqs}.

The full momentum-dependent SC Boltzmann equations are difficult to solve without simplifying approximations. One possibility is to integrate the SC Boltzmann equations into a set of moment equations, with CP-violating source terms induced by the SC force~\cite{Cline:1997vk,Cline:2000nw,Fromme:2006wx}. The current state of the art are the two-moment equations derived in~\cite{Cline:2020jre}, which are valid for arbitrary wall velocities. An alternative approach is to parametrize the non-equilibrium perturbations with some ansatz, or expand them in terms of some basis functions, and form a set of weighted integrals of the SC-equations to determine the coefficient functions in the expansion~\cite{Moore:1995si,Moore:1995ua,Laurent:2020gpg,Friedlander:2020tnq,Dorsch:2021ubz,Laurent:2022jrs}. In this paper we will study the first option, and generalize the moment equations of~\cite{Fromme:2006wx,Cline:2020jre} systematically to an arbitrary number of velocity moments. 

We will use the physical model introduced in~\cite{Cline:2017qpe} as a benchmark, and carefully study the dependence of the baryon asymmetry on the number of moments, and on extra truncation and factorization assumptions necessary to obtain a closure. We find that the final BAU shows little dependence on factorization choices, but is sensitive on the number of moments and the truncation choice. For large wall velocities the correction can be a factor of five as compared to the simplest two moment setup. For small wall velocities, with the usual truncation choices, no convincing asymptotic convergence takes place for large $n$. However, our new truncation scheme, based on setting the highest {\em variance} to zero, appears to give reasonably robust results for all wall velocities. Overall, the corrections we find are large enough to significantly affect the results from phenomenological parameter scans in various beyond standard model candidates for explaining the baryon asymmetry in the universe. 

This paper is organized as follows. In section~\cref{sect:BEqs} we review the derivation of the  semiclassical (SC) Boltzmann equation in the wall frame. In section~\cref{sec:moments} we derive a generic moment expansion from the SC-equations, including a discussion of truncation and factorization approximations needed for the closure. In section~\cref{sec:collision-terms} we introduce an approximation scheme that allows us to write the collision terms in a closed form in the moment expansion, and in section~\cref{sec:Full_Moment_eq_network} we summarise these results in a generic matrix equation with arbitrary number of moments and particle species. In section~\cref{sec:Baryon_asymmetry} we discuss the evaluation of Baryon asymmetry based on the chemical potentials obtained from moment equations, and in section~\cref{sec:bencmark-model} we introduce our benchmark model for the numerical analysis. In section~\cref{sec:numerical_results} we display our numerical results, and we conclude in section~\cref{sec:conclusions}.

%%%%%%%%%%%%%%%%%%%%%%%%%%%%%%%%%%%%%%%%%%%%%%%%%%%%%%%%%%%%%%%%%%%%%%%%%%%%%%%%%%%%%%%%%%%%%%%%%%%
%%%%%%%%%%%%%%%%%%%%%%%%%%%%%%%%%%%%%%%%%%%%%%%%%%%%%%%%%%%%%%%%%%%%%%%%%%%%%%%%%%%%%%%%%%%%%%%%%%%
%
\section{Boltzmann equation}
\label{sect:BEqs}
%
%%%%%%%%%%%%%%%%%%%%%%%%%%%%%%%%%%%%%%%%%%%%%%%%%%%%%%%%%%%%%%%%%%%%%%%%%%%%%%%%%%%%%%%%%%%%%%%%%%%
%%%%%%%%%%%%%%%%%%%%%%%%%%%%%%%%%%%%%%%%%%%%%%%%%%%%%%%%%%%%%%%%%%%%%%%%%%%%%%%%%%%%%%%%%%%%%%%%%%%

In this section, we review the semiclassical force method at the level of Boltzmann equations, following~\cite{Kainulainen:2001cn,Kainulainen:2002th,Cline:2017qpe,Cline:2020jre,Kainulainen:2021oqs}. We will be working in the rest frame of the wall, where the solutions are time independent. In this case, the full Boltzmann equation for the particle densities $f$ can be written as
\begin{equation}
   v_{h\pm}\partial_z f + F_{h\pm} \partial_{p_z} f_{h\pm} = {\cal C}_{h\pm}[f],
\label{eq:bolzmann_equation}
\end{equation}
where $h$ refers to helicity and $\pm$ to particles and antiparticles, $v_{h\pm}$ is the group velocity in the $z$-direction\footnote{
%
% FOOTNOTE BEGINS
%
Actually, the true group velocity is 
$v_{gh\pm} \equiv \partial_{k_z} \omega_{h\pm} = Z_{h\pm}v_{h\pm}$ and similarly, the actual semiclassical force is $-\partial_{z} \omega_{h\pm} = Z_{h\pm}F_{h\pm}$, where $Z_{h\pm}$ is the WKB-wave function renormalization factor~\cite{Kainulainen:2021oqs}. In practice $Z_{h\pm}$ is divided out and absorbed into the collision term, where it can eventually be neglected to the order we are working in the gradient expansion.}, 
%
% FOOTNOTE ENDS
%
$F_{h\pm}$ is the semiclassical force and ${\cal C}_{h\pm}[f]$ is a collision integral to be defined later. For a fermion with a $CP$-violating complex mass term $\hat m(z) = m(z)e^{i\gamma^5\theta(z)}$ these functions become~\cite{Cline:2017qpe,Cline:2020jre}:
\be
  v_{h\pm} = \frac{p_z}{\omega_{h\pm}} \qquad {\rm and} \qquad
  F_{h\pm} = -\frac{|m|^{2\prime}}{2\omega_{h\pm}} \pm s_h \frac{(|m|^2\theta')'}{2\omega_0\omega_{0z}},
\label{eq:vgF_2b}
\ee
to second order in the spatial derivatives denoted by $\partial_za = a^\prime$. To the same order in derivatives, the wall-frame conserved energy can be written as
\begin{equation}
\omega_{h\pm} \approx \omega_{0} \mp s_h \frac{|m|^2\theta'}{2\omega_{0}\omega_{0z}}.
\label{eq:omegaw}
\end{equation}
Here we defined $\omega_0 \equiv \sqrt{{\bm p}^2 + m^2}$ and $\omega_{0z} \equiv \sqrt{p_z^2 + m^2}$ in terms of a physical momentum variable ${\bm p}$. Finally, the spin factor $s_h$ is given by
\be
s_h = \langle p,h | S^{w}_z | p,h \rangle  
   = h\gamma_{||}\frac{p_z}{|{\bm p}|} \equiv h s_{\rm p},
\label{eq:shel}
\ee
where $h=\pm1$ is the helicity and $\gamma_{||} = \omega_0/\omega_{0z}$ is the boost factor to the ${\bm p}_{||}=0$-frame.

For bosonic fields there is no CP-violating force in the order we are working~\cite{Cline:2000nw}. On the other hand, there is a CP-even force acting on bosons, which to leading order in gradients is the same as for fermions. So equations~\cref{eq:vgF_2b,eq:omegaw} are also valid for bosons when one sets $s_h=0$ everywhere. This will be useful below, because our network of equations will also contain bosons interacting with fermions.

%%%%%%%%%%%%%%%%%%%%%%%%%%%%%%%%%%%%%%%%%%%%%%%%%%%%%%%%%%%%%%%%%%%%%%%%%%%%%%%%%%%%%%%%%%%%%%%%%%%
%
\subsection{Boltzmann equation for the perturbations}
%
%%%%%%%%%%%%%%%%%%%%%%%%%%%%%%%%%%%%%%%%%%%%%%%%%%%%%%%%%%%%%%%%%%%%%%%%%%%%%%%%%%%%%%%%%%%%%%%%%%%

The Boltzmann equation~\cref{eq:bolzmann_equation} is not very practical in its generic form. It is useful to separate the equilibrium part of the distribution from the out-of-equilibrium perturbations. To this end we define
\be
f_{h\pm} = f_\FD(\gamma_w(\omega_{h\pm}+v_w p_z)) + \Delta f_{h\pm},
\label{eq:fplupert}
\ee
where $\gamma_w = 1/\sqrt{1-v_w^2}$, $f_\FD$ is the equilibrium Fermi-Dirac distribution, and $v_w$ is the relative velocity of the plasma in front of the wall with respect to the wall.\footnote{
%
% FOOTNOTE BEGINGS
%
Note that this in general differs from the wall velocity with respect to the rest frame of the bubble centre, since the plasma is in motion around the wall. For strong transitions $v_w$ would also vary across the wall, but we shall ignore this effect here.} 
%
% FOOTNOTE ENDS
%
It is easy to see that $v_{h\pm}\partial_z f_\FD( \omega_{h\pm}) + F_{h\pm} \partial_{p_z}f_\FD(\omega_{h\pm})= 0$, so the equilibrium distribution satisfies the Liouville equation (equation~\cref{eq:bolzmann_equation} without a collision term) for $v_w=0$. However, in the case of a moving wall, the Liouville operator acting on the equilibrium distribution creates a source term for the perturbation $\Delta f_{h\pm}$. Inserting~\cref{eq:fplupert} into equation~\cref{eq:bolzmann_equation}, one finds:
\begin{equation}
v_{h\pm}\partial_z \Delta f_{h\pm} + F_{h\pm} \partial_{p_z} \Delta f_{h\pm} = - v_w\gamma_w F_{h\pm} (f_\FD^{h\pm})^\prime + {\cal C}_{h\pm}[f],
\label{eq:bolzmann_equationdel}
\end{equation}
where prime acting on $f$ denotes $\partial f/\partial (\gamma_w\omega)$. We further divide the perturbation into CP-even and CP-odd parts:
\begin{equation}
\Delta f_{h\pm} = \Delta f \pm \Delta f_h,
\label{eq:split-to-even-odd}
\end{equation}
where only the CP-odd perturbation $\Delta f_h$ depends on helicity. One can derive equations for $\Delta f$ and $\Delta f_h$ by taking the sum and the difference of the equation~\eqref{eq:bolzmann_equationdel}. These equations mix, but this occurs only at the third order or higher in gradients~\cite{Fromme:2006wx,Cline:2020jre}, so we can treat the equations independently. Expanding consistently to second order in gradients, one finds the CP-odd equation:
\begin{equation}
\frac{p_z}{\omega_0} \partial_z \Delta f_h - \frac{\m2p}{2\omega_0} \partial_{p_z} \Delta f_h = {\cal S}_h + {\cal C}_h[f],
\label{eq:bolzmann_equationdel2}
\end{equation}
where the collision term is ${\cal C}_h = ({\cal C}_{h+}-{\cal C}_{h-})/2$ and the CP-violating source is:
\be
{\cal S}_h 
=-v_w \gamma_w s_h \,\gamma_\PA \left[ \frac{(|m|^2\theta')'}{2 \omega_0^2}f'_{0w}
   - \frac{\m2p |m|^2\theta^\prime}{4 \omega_0^4}
	\left(f'_{0w} - \gamma_w \omega_0 f''_{0w}\right) \right] ,
\label{eq:CPodd_source}
\ee
where $f_{0w} \equiv f_\FD(\gamma_w(\omega_0 + v_w p_z))$. The CP-even perturbation $\Delta f$ satisfies a similar equation with the replacement ${\cal S}_h \rightarrow -\sfrac{1}{2}v_w\gamma_w(|m^2|'/\omega_0)f'_{0w}$. In this paper we will work at the level of vacuum dispersion relations. Thermal corrections to the semiclassical source have been computed in  ref.~\cite{Kainulainen:2021oqs}, but their implementation on the moment expansion is beyond the scope of this work.

%%%%%%%%%%%%%%%%%%%%%%%%%%%%%%%%%%%%%%%%%%%%%%%%%%%%%%%%%%%%%%%%%%%%%%%%%%%%%%%%%%%%%%%%%%%%%%%%%%%
%%%%%%%%%%%%%%%%%%%%%%%%%%%%%%%%%%%%%%%%%%%%%%%%%%%%%%%%%%%%%%%%%%%%%%%%%%%%%%%%%%%%%%%%%%%%%%%%%%%
%
\section{SC-moment equations}
\label{sec:moments}
%
%%%%%%%%%%%%%%%%%%%%%%%%%%%%%%%%%%%%%%%%%%%%%%%%%%%%%%%%%%%%%%%%%%%%%%%%%%%%%%%%%%%%%%%%%%%%%%%%%%%
%%%%%%%%%%%%%%%%%%%%%%%%%%%%%%%%%%%%%%%%%%%%%%%%%%%%%%%%%%%%%%%%%%%%%%%%%%%%%%%%%%%%%%%%%%%%%%%%%%%

We now set up a systematic moment expansion method for solving~\cref{eq:bolzmann_equationdel2}. It is useful to further separate the perturbation into a piece corresponding to a finite effective chemical potential and an additional momentum-dependent perturbation. We display the results explicitly only for the CP-odd perturbation, but identical results apply for the CP-even equations, only with a different source term. We thus define:
\begin{equation}
\Delta f_h \equiv - \mu_hf_{0w\pm}^\prime + \delta f_h,
\label{eq:split-mu-deltaf}
\end{equation}
with an additional integral constraint
\be
\int \frac{{\rm d}^3p}{(2\pi )^3} \delta f_h \equiv 0,
\label{eq:constraint}
\ee
which defines the chemical potential $\mu_h$ unambiguously. Inserting the split~\cref{eq:split-mu-deltaf} into equation \cref{eq:bolzmann_equationdel2}, the Liouville operator in the left hand side becomes
\begin{equation}
L[\mu_h,\delta f_h] \equiv  -\frac{p_z}{\omega_0}f_{0w}^\prime \,\partial_z\mu_h + v_w\gamma_w\frac{\m2p}{2\omega_0}f^{\prime\prime}_{0w} \mu_h  + \frac{p_z}{\omega_0} \partial_z \delta f_h - \frac{\m2p}{2\omega_0}\partial_{p_z}\delta f_h. 
\label{eq:Liouville}
\end{equation}
Integrating the~\cref{eq:bolzmann_equationdel2} over the spatial momenta weighted by $(p_z/\omega_0)^\ell$, with different $\ell$, one obtains a set of equations for $\mu_h$ and the velocity moments of the perturbation $\delta f_h$. To be precise, we define the integration over momenta by the average\footnote{
%
% FOOTNOTE BEGIN
%
Note that our normalization differs by one power of $T$ from that of~\cite{Cline:2020jre}.}: 
%
% FOOTNOTE END
%
\begin{equation}
\langle X \rangle \equiv \frac{1}{N_1} \int {\rm d}^3p \, X,
\label{eq:velocity-moment}
\end{equation}
with $N_1 = - 2\pi^3\gamma_w T^3/3$. The $\ell$'th moment equation then becomes:
\begin{align}
- \Big\langle \frac{p_z^{\ell+1}}{\omega_0^{\ell+1}} f'_{0w} \Big\rangle \partial_z \mu_h 
&+ \frac{1}{2}v_w\gamma_w \m2p \Big\langle \frac{p_z^\ell}{\omega_0^{\ell+1}}f^{\prime\prime}_{0w}\Big\rangle \mu_h
\nonumber \\
&+ \Big\langle \frac{p_z^{\ell+1}}{\omega_0^{\ell+1}}\partial_z\delta f_h\Big\rangle
 - \frac{1}{2}\m2p \Big\langle \frac{p_z^{\ell}}{\omega_0^{\ell+1}}\partial_{p_z}\delta f_h\Big\rangle
    =  \Big\langle \frac{p_z^{\ell}}{\omega_0^{\ell}} \big( S^w_{h} + C^w_{h} \big) \Big\rangle.
\label{eq:pregenericmomeq}
\end{align}
One can further simplify this equation by using the following identity:
\be
\Big\langle \frac{p_z^{\ell+1}}{\omega_0^{\ell+1}}\partial_z\delta f_h\Big\rangle
 - \frac{1}{2}\m2p \Big\langle \frac{p_z^{\ell}}{\omega_0^{\ell+1}}\partial_{p_z}\delta f_h\Big\rangle
 = \partial_z \Big\langle \frac{p_z^{\ell+1}}{\omega_0^{\ell+1}} \delta f_h \Big\rangle + \frac{\ell}{2}\m2p  \Big\langle \frac{p_z^{\ell-1}}{\omega_0^{\ell+1}}\delta f_h\Big\rangle.
 \label{eq:intermedstep}
\ee
Next we define the dimensionless $\ell$'th velocity moment of the distribution $\delta f$ along with a dimensionless chemical potential and dimensionless mass variable as follows:
\be
u_{h,\ell} \equiv \Big\langle \frac{p_z^\ell}{\omega_0^\ell} \delta f_h \Big\rangle,  
\qquad \xi_i \equiv \frac{\mu_i}{T} \quad {\rm and} \quad x \equiv \frac{m}{T}.
\label{eq:velocity-momentb}
\ee
We further divide equation~\cref{eq:pregenericmomeq} by $T$, after which we can write the $\ell$'th moment equation in a dimensionless form where all dimensionful quantities are measured in units of temperature:
\begin{equation}
\boxed{
\phantom{\Big[} -D_{\ell + 1} \xi_h' + u_{h,\ell+1}^\prime  + v_w\gamma_w |x|^{2\prime} Q_\ell\xi_h +\ell |x|^{2\prime} \bar R u_{h,\ell} 
    =  \hat{\cal S}^w_{h,\ell} + \hat{\cal C}^w_{h\ell}. \phantom{\Big]} }
\label{eq:genericmomeq}
\end{equation}
Here the prime denotes the dimensionless spatial derivative $' \equiv \partial/\partial(zT)$ and $\ell$ runs from 0 to $n-1$, giving $n$ coupled moment equations. The dimensionless coefficient functions $D_\ell$ and $Q_\ell^h$ are given by:
\begin{equation}
D_\ell \equiv T \Big\langle \frac{p_z^\ell}{\omega_0^\ell} \!f_{0w}^\prime \Big\rangle, \quad 
Q_\ell \equiv T^3 \Big\langle \frac{p_z^\ell}{2\omega_0^{\ell+1}} \!f_{0w}^{\prime\prime} \Big\rangle,
\quad{\rm and}\quad 
\bar R u_{h,\ell} \equiv T^2\Big\langle \frac{p_z^{\ell-1}}{2\omega_0^{\ell + 1}}\delta f_h \Big\rangle,
\label{eq:D-and-Q}
\end{equation}
and the thermally corrected dimensionless source function in the wall frame is
\begin{equation}
\hat{\cal S}^w_{h,\ell} = -v_w \gamma_w \Big[ (|x|^2\theta^\prime)^\prime Q^{8o}_{h,\ell} - \x2p |x|^2\theta^\prime 
Q^{9o}_{h,\ell} \Big],
\label{eq:sterms}
\end{equation}
where
\begin{align}
 Q^{8o}_{h,\ell} &\equiv T^3\Big\langle \frac{s_hp_z^\ell}{2\omega_0^{\ell+1} \omega_{0z}} f_{0w}^\prime \Big\rangle
\label{eq:Q8} 
\\
Q^{9o}_{h,\ell} &\equiv T^5\Big\langle \frac{s_hp_z^\ell}{4\omega_0^{\ell+2}\omega_{0z}} 
     \Big( \frac{1}{\omega_0}f_{0w}^\prime -\gamma_w f_{0w}^{\prime\prime} \Big) \!\Big\rangle.
\label{eq:Q9} 
\end{align}
Finally, the temperature-scaled, dimensionless collision term is given by 
\begin{equation}
\hat {\cal C}^w_{h,\ell} \equiv \frac{1}{T}\Big\langle \frac{p_z^\ell}{{\omega_0}^\ell} C_{h}[f]\Big\rangle.
\label{eq:scaled_C}
\end{equation}
The reduction of the collision term will require a more detailed discussion and further approximations. We shall return to this question in section~\cref{sec:collision-terms} below.

%%%%%%%%%%%%%%%%%%%%%%%%%%%%%%%%%%%%%%%%%%%%%%%%%%%%%%%%%%%%%%%%%%%%%%%%%%%%%%%%%%%%%%%%%%%%%%%%%%%
%
\subsection{Factorization and truncation}
%
%%%%%%%%%%%%%%%%%%%%%%%%%%%%%%%%%%%%%%%%%%%%%%%%%%%%%%%%%%%%%%%%%%%%%%%%%%%%%%%%%%%%%%%%%%%%%%%%%%%

We have not yet defined the $\bar R$-term in~\cref{eq:genericmomeq}, which does not automatically correspond to any velocity moment function. To do this we slightly generalize the approach first suggested in~\cite{Fromme:2006wx} and define the following \emph{factorization} rule:
\begin{equation}
\bar R u_{h,\ell} \equiv T^2\Big\langle \frac{p_z^{\ell-1}}{2\omega_0^{\ell + 1}}\delta f_h \Big\rangle \rightarrow \Big[\frac{T^2}{2p_z\omega_0}\Big] u_{h,\ell},
\label{eq:rbar}
\end{equation}
where
\begin{equation}
    [{\cal X}]  \equiv \frac{1}{N_0} \int d^{\,3}p\, {\cal X} f_{0w}.
\label{eq:factorize}
\end{equation}
Here $N_0 = \int d^{\,3}p\, f_{0w} = \gamma_w \int d^{\,3}p\, f_{0} \equiv  \gamma_w \hat N_0$ is defined in terms of the massive plasma frame distribution function $f_0$ of the particle under consideration. Using~\cref{eq:factorize} one can compute $\bar R$ explicitly~\cite{Cline:2020jre}
\be
\bar R = \frac{\pi T^2}{\gamma_w^2 \hat N_0} \int_m^\infty {\rm d}\omega 
   \ln\left|\frac{p-v_w \omega}{p + v_w\omega}\right| f_0(\omega).
\label{eq:barR}
\ee

Two quantities in equation~\cref{eq:genericmomeq} still need to be specified. Obviously, we need the moments over the collision integral, which we will define in section~\cref{sec:collision-terms}. In addition, the highest velocity moment derivative $u'_{h,n}$ appearing in the equation with $\ell = n-1$, needs to be specified, as we do not have enough equations to determine it. For this we need a {\em truncation} scheme which relates $u_{h,n}$ to the lower moments. There are several different ways to do this.  First, one could again use the factorization rule:
\be
u_{h,n} = \Big\langle\!\left( \frac{p_z}{\omega_0} \right)^{n} \delta f \Big\rangle \rightarrow \Big[\frac{p_z}{\omega_0}\Big] u_{h,n-1} \equiv R u_{h,n-1}.
\label{eq:trunction-rule}
\ee
In this definition $R$ is just the expectation value of the fluid velocity in the wall frame, which is exactly given by $R = -v_w$. This is the truncation choice used in the early literature~\cite{Fromme:2006wx,Cline:2020jre}. 
Alternatively, one could just put the last moment to zero, setting $R=0$, or equal to the previous moment, setting $R=1$. In all these cases with a constant $R$, the truncation condition for $u'_{h,n}$ is just
\be
u'_{h,n} =  R u'_{h,n-1}.
\label{eq:trunction-condition}
\ee

%%%%%%%%%%%%%%%%%%%%%%%%%%%%%%%%%%%%%%%%%%%%%%%%%%%%%%%%%%%%%%%%%%%%%%%%%%%%%%%%%%%%%%%%%%%%%%%%%%%
\paragraph{Variance truncation.}
%%%%%%%%%%%%%%%%%%%%%%%%%%%%%%%%%%%%%%%%%%%%%%%%%%%%%%%%%%%%%%%%%%%%%%%%%%%%%%%%%%%%%%%%%%%%%%%%%%%

Finally, we propose an alternative truncation scheme, where we set the $n$'th {\em variance} to zero, which here can be interpreted as implying:
\be
\Big\langle\! \Big( \frac{p_z}{\omega_0} - u_{h,1} \Big)^{n} \delta f \Big\rangle = (-1)^{n+1} u_1^n.
\label{eq:R}
\ee
This condition defines the $n$'th moment as a particular series of all lower moment functions. The term in the {\em r.h.s.}~of~\cref{{eq:R}} is optional and it was inserted because the zeroth moment of $\delta f$ is vanishing. We have checked that the variance condition is not sensitive to keeping this term. At any rate, the condition~\cref{eq:R} can be written as a generalization of~\cref{eq:trunction-condition}:
\be
u'_{h,n} = \sum_{i=1}^{n-1}R_i u'_{h,i},
\label{eq:general_trunction-condition}
\ee
where the coefficients $R_i$:s now are functions of $u_i$:
\begin{align}
R_1 & = (-1)^n n(n-1)u_1^{n-1} + \sum_{k=2}^{n-1}(-1)^{n-k-1}\sfrac{n!}{(n-k-1)!k!}u_1^{n-k-1}u_k
\nonumber \\
R_i & = (-1)^{n-i-1}\sfrac{n!}{(n-i)!i!}u_1^{n-i}, \qquad {i = 2,...,n-1}.
\end{align}
Dropping the {\em r.h.s.}~in equation~\cref{eq:R} would simply change $n(n-1)\rightarrow n^2$ in the first term in the expression for $R_1$. 

Different choices for factorization and truncation conditions should not affect the solutions significantly. We have checked that shifting from the factorization rule~\cref{eq:barR} to setting $\bar R \equiv 0$ indeed changes the final baryon asymmetry only at a per cent level. Truncation dependence, however, turns out to be more severe; we will study this issue in detail in section~\cref{sec:truncation_dependence}. We shall see that the variance truncation scheme gives more robust predictions for the baryon asymmetry than the constant $R$-truncation schemes do.

%%%%%%%%%%%%%%%%%%%%%%%%%%%%%%%%%%%%%%%%%%%%%%%%%%%%%%%%%%%%%%%%%%%%%%%%%%%%%%%%%%%%%%%%%%%%%%%%%%%
%
\subsection{Inverted moment equations}
\label{sec:inverting_moment_equation}
%
%%%%%%%%%%%%%%%%%%%%%%%%%%%%%%%%%%%%%%%%%%%%%%%%%%%%%%%%%%%%%%%%%%%%%%%%%%%%%%%%%%%%%%%%%%%%%%%%%%%

Assembling the previous results, and including the collision and source terms, the fluid equations can now be presented in full detail. Defining a vector $w_h = (\xi_h,u_1,...,u_{n-1})^T$, the general form of the moment equations for one particle species may be expressed as a matrix equation
\be
\hat{\cal A} w_h' +  \hat{\cal B}[w_h] = \! \hat{\cal S}^w_h + \hat{\cal C}^w_h,
\label{WKBeqs}
\ee
where $\hat{\cal A}$ is an $n\times n$-dimensional matrix and $\hat{\cal B}[w_h]$ an $n$-dimensional column vector given by: 
\be
  \hat{\cal A} 
  = \left( \!\! \begin{array}{ccccc}   -D_1     &     1      &  \cdots   &    0      &    0    \\
                                       -D_2     &     0      &  \cdots   &    0      &    0    \\
	                               \vdots    &    \vdots  &  \ddots   &  \vdots   &  \vdots \\
	                              -D_{n-1}    &     0      &  \cdots   &    0      &    1    \\
	                               -D_n      &     R_1    &  \cdots   &  R_{n-2}  &  R_{n-1}   
	              \end{array}\!\right)
\,,\quad 
   \hat{\cal B}_\ell[w_h] = \frac{1}{2}\x2p (v_w\gamma_w w_{h,0}Q_\ell + \ell \bar R_1 w_{h,\ell}).
\label{eq:AB}
\ee
Likewise $\hat{\cal S}^w_h = (\hat{\cal S}^w_{h1},...,\hat{\cal S}^w_{h,n})^T$ and similarly for the $\hat{\cal C}^w_h$ vector. This form of moment equations is generic to both CP-even and CP-odd sectors, which only differ by the form of the source terms. Matrix $A$ is easily inverted, giving the equations in the simple form:
\be
w_h'  = \hat{\cal A}^{-1} (\hat{\cal S}^w_h + \hat{\cal C}^w_h - \hat{\cal B}[w_h]),%\;\;} 
\label{inverse-WKBeqs}
\ee
where the inverse matrix is:
\def\mcDn{{\mathcal{D}_n}}
\be
  \hat{\cal A}^{-1} = \frac{1}{\mcDn}
    \left( \!\! \begin{array}{cccccc}  
    R_1           &    R_2         & \cdots &   R_{n-1}         &    -1      \\
    R_1D_1        &    R_2D_1      & \cdots &   R_{n-1}D_1      &    -D_1     \\
    \vdots        &    \vdots      & \ddots &   \vdots          &    \vdots   \\
    R_1 D_{n-1}   &    R_2D_{n-1}  & \cdots &   R_{n-1}D_{n-1}  &    -D_{n-1}
    \end{array}\!\right)
    +\left( \!\! \begin{array}{cccccc}  
    0        &    0       & \cdots &   0     \\
    1        &    0       & \cdots &   0     \\
    \vdots   &    \ddots  & \ddots &   0     \\
    0        &    \cdots  &   1    &   0
    \end{array}\!\right) .
\label{inverse-WKBeqs}
\ee
Here the determinant factor $\mathcal{D}_n$ is given by 
\begin{equation}
\mathcal{D}_n \equiv (-1)^n \det(\hat{\cal A}) = D_n - \sum_{i=1}^{n-1}R_iD_i.
\label{eq:detA}
\end{equation}
Both the matrix ${\cal A}$ and its inverse $\hat{\cal A}^{-1}$ were written such that they are consistent with the general truncation rule~\cref{eq:general_trunction-condition}. The determinant factor $\mathcal{D}_n$ is a function of the wall-velocity as well as and $z$ and $n$. It is important that $\mathcal{D}_n$ is always nonzero so that the inverse $\hat{\mathcal{A}}^{-1}$ exists and this is indeed always the case in our calculations.

%==================================================================================================
%==================================================================================================

\begin{figure}
\centering
\includegraphics[width=0.83\textwidth]{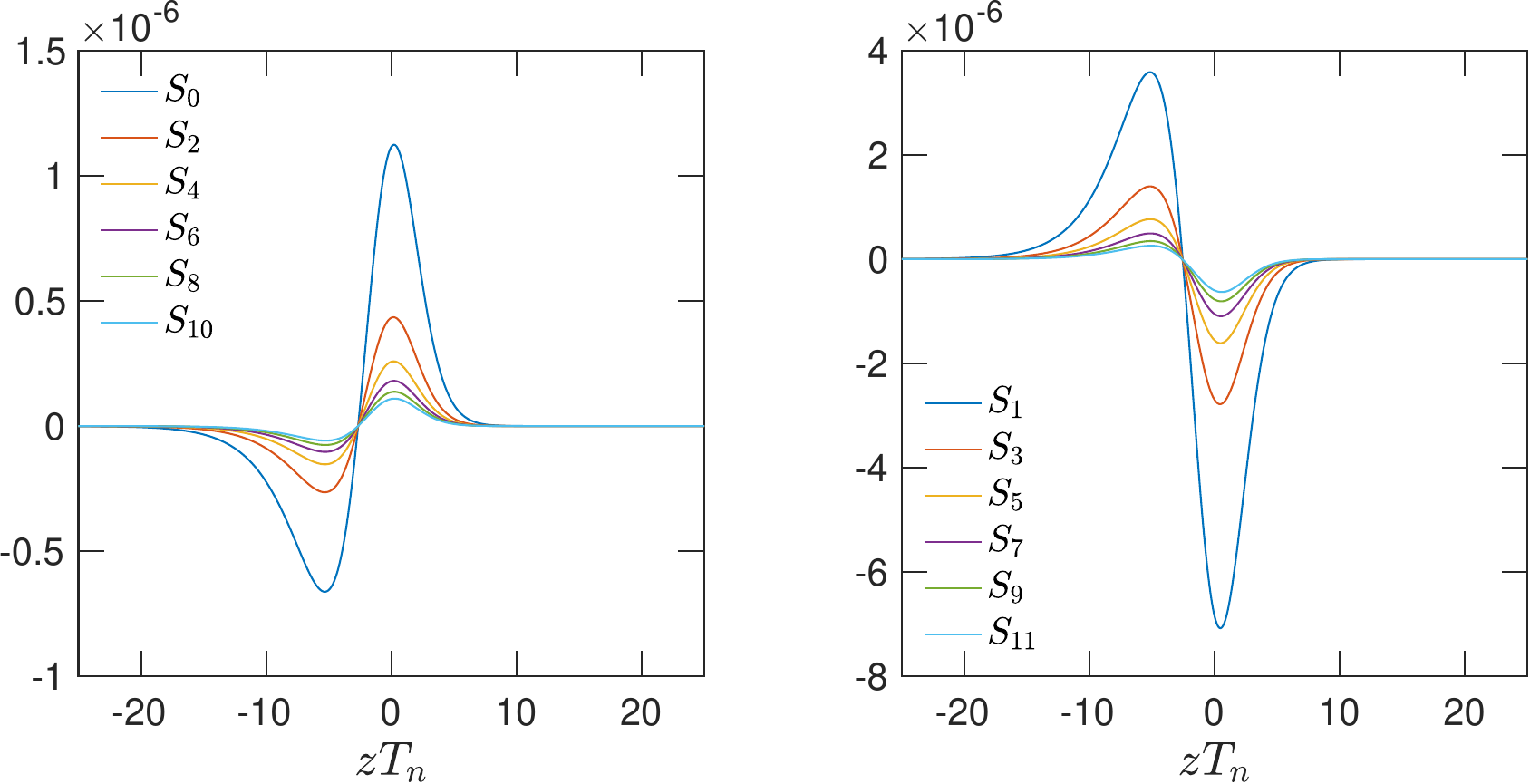}
\caption{Shown are few first source functions $S^w_{h,\ell}$, defined in~\cref{eq:sterms} for $v_w=0.1$. Left panel shows the sources for even $\ell = 0,2,4,6,8,10$ and right panel the odd $\ell = 1,3,5,7,9,11$. The mass function is defined below in equation~\cref{mteq} and we used $v_w=0.1$ and the benchmark parameters given in equation~\cref{eq:fiducial_model}.}
\label{fig:sources}
\end{figure}
%==================================================================================================
%==================================================================================================

Equations~\cref{inverse-WKBeqs} depend on a large number of external functions: $D_{\ell}$, $Q_{\ell}$, $R$, $\bar R$, $Q^{e}_{\ell}$, $Q^{8o}_{\ell}$ and $Q^{9o}_{\ell}$, which in general depend on two variables, the wall velocity $v_w$ and the dimensionless ratio $x = |m|/T$, and most of which can only be evaluated numerically. However, all these functions are universal and we have evaluated them for a large number of $n$ in a grid in $x$ and $v_w$. Our numerical package automatically loads the data and creates spline-fits for all external functions needed, which can then be used as regular functions. We give explicit forms for all external functions in terms of a generic integral function in the Appendix~\cref{sec:app_explicit_forms}.

%%%%%%%%%%%%%%%%%%%%%%%%%%%%%%%%%%%%%%%%%%%%%%%%%%%%%%%%%%%%%%%%%%%%%%%%%%%%%%%%%%%%%%%%%%%%%%%%%%%
%
\subsection{Sources and expected convergence of the expansion}
\label{sec:source}
%
%%%%%%%%%%%%%%%%%%%%%%%%%%%%%%%%%%%%%%%%%%%%%%%%%%%%%%%%%%%%%%%%%%%%%%%%%%%%%%%%%%%%%%%%%%%%%%%%%%%

To finish this section, let us briefly consider the typical form and the size of the source terms in the fluid equations. We plot the source terms~\cref{eq:sterms} for the first few moments in figure~\cref{fig:sources}. Note how the sources get smaller with increasing $\ell$. This is expected, since on average $p_z/\omega \ll 1$, and so adding higher and higher powers of this factor should make the source progressively smaller. Note also that the sources for odd $\ell$--moments are systematically larger than those for even $\ell$.  This follows from the symmetry properties of functions $Q^{8o}_{\ell}$ and $Q^{9o}_{\ell}$. Indeed, in the limit $v_w=0$ both functions are, for a given $h$, symmetric for odd $\ell$ and anti-symmetric for even $\ell$ under reflection $p_z\rightarrow -p_z$ (for a fixed $h$, $s_h (- p_z) = - s_h (p_z)$). Thus the even $\ell$--sources suffer from additional velocity suppression in comparison to the odd $\ell$--sources. Overall, these results show that the higher moment equations are sourced more weakly than are the lower moment equations.

%%%%%%%%%%%%%%%%%%%%%%%%%%%%%%%%%%%%%%%%%%%%%%%%%%%%%%%%%%%%%%%%%%%%%%%%%%%%%%%%%%%%%%%%%%%%%%%%%%%
%%%%%%%%%%%%%%%%%%%%%%%%%%%%%%%%%%%%%%%%%%%%%%%%%%%%%%%%%%%%%%%%%%%%%%%%%%%%%%%%%%%%%%%%%%%%%%%%%%%
%
\section{Collision integrals in the moment expansion}
\label{sec:collision-terms}
%
%%%%%%%%%%%%%%%%%%%%%%%%%%%%%%%%%%%%%%%%%%%%%%%%%%%%%%%%%%%%%%%%%%%%%%%%%%%%%%%%%%%%%%%%%%%%%%%%%%%
%%%%%%%%%%%%%%%%%%%%%%%%%%%%%%%%%%%%%%%%%%%%%%%%%%%%%%%%%%%%%%%%%%%%%%%%%%%%%%%%%%%%%%%%%%%%%%%%%%%

In this section we specify the collision integrals for the moment expansion. Collision terms are intrinsically model dependent, and their momentum dependence is not expressible as an expansion in powers of $p_{z}/\omega_0$. Reducing them to a form suitable for the moment expansion method requires several approximations, which we explain in more detail in a companion paper~\cite{KimmoNiyati2}. Here we will give sufficient details to justify the {\em form} of the collision integrals, but we shall not actually compute the rate-functions appearing in the final expressions. Moreover, we will be computing all collision integrals to the lowest order in gradients, which is consistent to the overall order we are working in the gradient expansion~\cite{Kainulainen:2021oqs}. 

%%%%%%%%%%%%%%%%%%%%%%%%%%%%%%%%%%%%%%%%%%%%%%%%%%%%%%%%%%%%%%%%%%%%%%%%%%%%%%%%%%%%%%%%%%%%%%%%%%%
\paragraph{Decay processes}
%%%%%%%%%%%%%%%%%%%%%%%%%%%%%%%%%%%%%%%%%%%%%%%%%%%%%%%%%%%%%%%%%%%%%%%%%%%%%%%%%%%%%%%%%%%%%%%%%%%

After a series of approximations following~\cite{Cline:2000nw} and given in more detail in~\cite{KimmoNiyati2}, we can write the dimensionless collision term $\hat C \equiv C/T$ for a species $i$, involved in a decay and inverse decay process $1 \leftrightarrow 23$ as follows:
\begin{equation}
    \hat C_{{\rm dec},i}(\xi,\delta f;p_i) \approx - c_i (\xi_1 - \xi_3 - \xi_4 ) 
    f^\MB_{0i}(p_i)\kappa_i \hat \Gamma_{\rm dec} -\delta f_i(p_i) \kappa_i \hat \Gamma_{\rm dec},
\label{eq:Cdec-2b}
\end{equation}
where $c_1=1$ and $c_{3,4}=-1$ and $\kappa_i = |N_1|/n^\MB_i$. Moreover, $\hat \Gamma_{\rm dec} = \langle (|m_1|/\omega_{10})f^\MB_{0i}\rangle (\Gamma_{\rm dec}^{\rm rf}/T)$ where $\Gamma_{\rm dec}^{\rm rf}$ is the usual rest frame decay rate. Even with the many approximations made, equation~\cref{eq:Cdec-2b} retains correct main features: the $\kappa_i$-factors ensure that the detailed balance holds and the collision term damps perturbations with the appropriate boosted decay rate. 

%%%%%%%%%%%%%%%%%%%%%%%%%%%%%%%%%%%%%%%%%%%%%%%%%%%%%%%%%%%%%%%%%%%%%%%%%%%%%%%%%%%%%%%%%%%%%%%%%%%
\paragraph{Scattering processes}
%%%%%%%%%%%%%%%%%%%%%%%%%%%%%%%%%%%%%%%%%%%%%%%%%%%%%%%%%%%%%%%%%%%%%%%%%%%%%%%%%%%%%%%%%%%%%%%%%%%

One can similarly reduce the elastic and inelastic scattering terms to an equally simple form containing chemical potentials $\xi_i$ and the perturbations $\delta f_i$. The collision integral for a general \textit{inelastic} $2\rightarrow 2$ scattering processes can be written as:
\begin{equation}
    \hat C_{{\rm inel},i}(p_i) \approx -\Big(\sum_j \nolimits s_{ij} \xi_j\Big) 
    f^\MB_{0i}(p_i)\kappa_i \hat \Gamma_{\rm inel} -\delta f_i(p_i) \kappa_i \hat \Gamma_{\rm inel},
\label{eq:Cinel-2}
\end{equation}
where $s_{ij}=1$ (-1) if the particle $j$ is in the initial (final) state of the collision integral for the species $i$. In \textit{elastic} channels the only difference is that the chemical potential term cancels and one finds:
\begin{equation}
  \hat C_{{\rm el},i}(p_i) \;\approx\; - \delta f_i (p) \kappa_i \hat \Gamma_{\rm el}.
  \label{eq:elastic-rates}
\end{equation}
Adding any number of channels necessary for a given equation is straightforward addition task using the generic results~\cref{eq:Cinel-2,eq:elastic-rates}. In these equations the only model dependent quantity is the effective dimensionless interaction rate $\hat \Gamma_{\rm X} = \big\langle f_{0i} \Gamma_{\rm X}(p_1)\big\rangle/T$, where X refers to the channel in question. In the chiral limit (massless particles) and in the Maxwell-Boltzmann approximation for the distribution functions $f_{0i}$ this function can be generically written as~\cite{KimmoNiyati2}:
\begin{equation}
\hat \Gamma_{\rm X} = \frac{1}{4\pi |N_1|} 
\int_{s_{\rm min}}^\infty {\rm d}s \, \frac{1}{\sqrt{s}}\lambda(s,m_1^2,m_2^2)
 K_1\Big(\frac{\sqrt{s}}{T}\Big)\sigma_{\rm X} (s) \,,
\label{eq:Gamma-as-Mollerintegral}
\end{equation}
where $|N_1| = 2\pi^3\gamma_w T^3/3$, $K_1(x)$ is the modified Bessel functions of the second kind, $\sigma_{\rm X}(s)$ is the CM-frame cross section and $s_{\rm min}$ is the kinematic threshold for the process $1+2\rightarrow 3+4$. Note that this expression would not be adequate to compute the collision integrals in the helicity basis for massive particles, because then the boost to the CM-frame performed to obtain~\cref{eq:Gamma-as-Mollerintegral} is not allowed. If such accuracy was required, the appropriate rate could be computed in small $v_w$-limit as a 5-dimensional integral over the phase space using methods developed in~\cite{Hannestad:1995rs,Ala-Mattinen:2022nuj}. 

%%%%%%%%%%%%%%%%%%%%%%%%%%%%%%%%%%%%%%%%%%%%%%%%%%%%%%%%%%%%%%%%%%%%%%%%%%%%%%%%%%%%%%%%%%%%%%%%%%%
\paragraph{Strong sphaleron rate}
%%%%%%%%%%%%%%%%%%%%%%%%%%%%%%%%%%%%%%%%%%%%%%%%%%%%%%%%%%%%%%%%%%%%%%%%%%%%%%%%%%%%%%%%%%%%%%%%%%%

The strong sphaleron rate (SSR) is a nonperturbative process, which causes equilibration of chirality in the SU(3)-sector of the theory. It is usually included at the integrated level, where it contributes a term in the first moment equation:
\begin{equation}
\hat C_{\rm SS}[\xi] = \hat \Gamma_{\rm SS}\sum_q \nolimits (\xi_{q_\L} \!-\!\xi_{q_\R}),
 \label{eq:SSR1}
\end{equation}
where the sum runs over all quark flavours. The rate $\Gamma_{\rm SS}$ is obtained from lattice simulations which measure the diffusion of the chiral Chern-Simons (CS) number~\cite{Moore:1997im,Moore:2010jd}, and offer little clue of its momentum dependence. Assigning higher moments to this rate is thus an ambiguous process, except that we know that CS-number diffusion involves only deep infrared physics, which should strongly suppress its contributions beyond the lowest moment. Following this idea we will include the strong sphaleron rate only to the lowest moment equation, but we have checked that the results are not very sensitive to this choice.

%%%%%%%%%%%%%%%%%%%%%%%%%%%%%%%%%%%%%%%%%%%%%%%%%%%%%%%%%%%%%%%%%%%%%%%%%%%%%%%%%%%%%%%%%%%%%%%%%%%
%
\subsection{General collision term}
\label{sec:general_collision_term}
%
%%%%%%%%%%%%%%%%%%%%%%%%%%%%%%%%%%%%%%%%%%%%%%%%%%%%%%%%%%%%%%%%%%%%%%%%%%%%%%%%%%%%%%%%%%%%%%%%%%%

Combining the results from the previous section we can write the collision integral in the $\ell$'th moment equation arising from inelastic and elastic processes for the species $a$ as follows:
\begin{equation}
  \hat C_{w,a}^{\ell} =  {\bar C}_{w,a} K^a_\ell - u_{a\ell} \kappa_a \hat \Gamma^a_{\rm tot},
\label{eq:final-collisionC}
\end{equation}
where the collision term from decays and $2\rightarrow 2$ scatterings is
\begin{align}
  \bar C_{w,a} = & \sum_{c,d} 
                 (\xi_a-\xi_c-\xi_d) \hat \Gamma^{{\rm dec}}_{a\rightarrow cd}
               + \sum_{b,c,d} 
                 (\xi_a + \xi_b-\xi_c-\xi_d) \hat \Gamma_{ab\rightarrow cd}.
\label{eq:final-collision_comp}
\end{align}
\vskip-0.2cm
\noindent The rate $\bar\Gamma_{ab\rightarrow cd}$ was defined in~\cref{eq:Gamma-as-Mollerintegral} and $\hat \Gamma^a_{\rm tot} = \Gamma_{\rm tot}/T$ is the total interaction rate with:
\begin{equation}
\hat \Gamma^a_{\rm tot} = \sum \big( \, \hat \Gamma^a_{\rm el} + \hat \Gamma_{ab\rightarrow cd} + \hat \Gamma^{\rm dec}_{a\rightarrow cd} \, \big),
\label{eq:gammatot_general}
\end{equation}
\noindent where the sum runs over all open channels. Finally the kinematic integral $K^a_\ell$ was defined as:
\begin{equation}
K^a_\ell \equiv - \kappa_a\Big\langle \frac{p_z^\ell}{\omega_{0a}^\ell} \!f^a_{0w} \Big\rangle 
= \Big[\frac{p_z^\ell}{\omega_{0a}^\ell}\Big]_a.
\label{eq:K-function}
\end{equation}
In particular then $K_0^a =1$ and $K_1^a = -v_w$ for all species. Moving from species $a$ to the equation for another species $b$, one has to replace $K_\ell^a\rightarrow K_\ell^b$ and switch the sign in chemical potential terms appropriately, and compute the elastic collision term $\kappa_b\hat \Gamma^b_{\rm tot}$.  Finally, for all quarks in the network one should add the strong-sphaleron contribution $ \hat C_{\mathrm{SS}\ell}^{a} =  \mp \hat\Gamma_{\rm SS}[\xi]$, to the lowest moment equation, where $+1$ corresponds to left- and $-1$ to right-chiral fields.

%%%%%%%%%%%%%%%%%%%%%%%%%%%%%%%%%%%%%%%%%%%%%%%%%%%%%%%%%%%%%%%%%%%%%%%%%%%%%%%%%%%%%%%%%%%%%%
%%%%%%%%%%%%%%%%%%%%%%%%%%%%%%%%%%%%%%%%%%%%%%%%%%%%%%%%%%%%%%%%%%%%%%%%%%%%%%%%%%%%%%%%%%%%%%
%
\section{Moment equation network coupling different species}
\label{sec:Full_Moment_eq_network}
%
%%%%%%%%%%%%%%%%%%%%%%%%%%%%%%%%%%%%%%%%%%%%%%%%%%%%%%%%%%%%%%%%%%%%%%%%%%%%%%%%%%%%%%%%%%%%%%
%%%%%%%%%%%%%%%%%%%%%%%%%%%%%%%%%%%%%%%%%%%%%%%%%%%%%%%%%%%%%%%%%%%%%%%%%%%%%%%%%%%%%%%%%%%%%%

So far we have only written the moment equations for a single species, but extension to an arbitrary number of species $N$ is straightforward. As explained in section~\cref{sec:inverting_moment_equation}, each particle species is described by an $n$-vector $w_a = (\xi_a,u_{a1},...,u_{a(n-1)})^T$ in the moment expansion. For a system of equations containing $N$ interacting species, these vectors can be combined into an $Nn$-vector $\mathcal{W}^T=(w_1^T,w_2^T,,...,w_N^T)$, which obeys an equation of the form 
\be
  \hat {\cal A} \mathcal{W}^{\,\prime}  =  \hat{\cal S} + \hat{\cal C}[\mathcal{W}]  - \hat{\cal B}[\mathcal{W}],
\label{eq:deqs}
\ee
where $\hat{\cal A} = {\rm diag}(\hat{\cal A}_1, \hat{\cal A}_2,...,\hat{\cal A}_N)$ is a block-diagonal matrix with $n$-dimensional blocks defined in~\cref{WKBeqs}. The source term $\hat{\cal S}={\rm diag}(\hat{\cal S}_1, \hat{\cal S}_2,...,\hat{\cal S}_N)$, is obviously an $Nn$-vector, whose elements are given by~\cref{eq:sterms} and while interactions terms in the collision integrals mix different particle species, and individual elements in $\hat{\cal B}_i$ mix moments within a given species, both $\hat{\cal C}[\mathcal{W}]$ and $\hat{\cal C}[\mathcal{W}]$ can still be written as $Nn$-vectors. Equation~\cref{eq:deqs} then has a  block-diagonal structure and it can be easily inverted to yield
\begin{equation}
 \mathcal{W}^{\,\prime} = \hat{\cal A}^{-1}\big( \hat{\cal S} + \hat{\cal C}[\mathcal{W}] - \hat{\cal B}[\mathcal{W}] \big),
 \label{eq:final_eq}
\end{equation}
where $\hat{\cal A}^{-1} = {\rm diag}(\hat{\cal A}_1^{-1}, \hat{\cal A}_1^{-1}, ...,\hat{\cal A}_N^{-1})$, with the individual terms $\hat{\cal A}_a^{-1}$ as defined in~\cref{inverse-WKBeqs}. This system can be solved using \eg~the relaxation method for any system of interest. Of course, both $\hat{\cal B} [\mathcal{W}]$ and $\hat{\cal C}[\mathcal{W}]$ could be written as matrices acting on $\mathcal{W}$, but it is easier to treat them numerically as functions of $\mathcal{W}$ that return $Nn$-vectors as was assumed above%
\footnote{In section~\cref{sec:eigenmode_analysis} below, where we discuss the convergence of the solutions at boundaries, we will take the different point of view and write $\hat{\cal C}[\mathcal{W}] = \hat\Gamma \mathcal{W}$, where $\hat\Gamma =\Gamma/T$ is an $(Nn)\times (Nn)$ matrix.}.  

Based on our results in the previous section, the generic element in the $\hat{\cal C}[\mathcal{W}]$-vector, the $\ell$'th moment of the collision integral in the equation for the species $a$, can be written as
\begin{equation}
\hat{\cal C}[\mathcal{W}]_{a\ell} = 
{\bar C}_a K_\ell^a - u_{a\ell}\kappa_a \hat\Gamma^a_{\rm tot} 
                \mp \delta_{0,\ell} \hat C^a_{\rm SS}[\xi],
\label{eq:general-collision-integral}
\end{equation}
where $\smash{{\bar C}_a}$ contains the decay and scattering contributions which, along with the $\smash{\hat\Gamma^a_{\rm tot}=\hat\Gamma^a_{\rm tot}/T}$, are the only model dependent quantities on the problem. In our benchmark model we will account for the top and bottom Yukawa interactions, the $W$ boson interactions that tend to equalize $\mu_i$'s within doublets and the helicity flip interactions that damp the helicity asymmetry $\mu_\tL \!-\! \mu_\tR$. In addition, the gauge interactions generate a Higgs chemical potential damping term in the broken phase. 

This completes our formal derivation of the moment equations for a generic system. Before moving to the explicit calculations in the benchmark model, we shall discuss the derivation of baryon asymmetry.

%%%%%%%%%%%%%%%%%%%%%%%%%%%%%%%%%%%%%%%%%%%%%%%%%%%%%%%%%%%%%%%%%%%%%%%%%%%%%%%%%%%%%%%%%%%%%%%%%%%
%%%%%%%%%%%%%%%%%%%%%%%%%%%%%%%%%%%%%%%%%%%%%%%%%%%%%%%%%%%%%%%%%%%%%%%%%%%%%%%%%%%%%%%%%%%%%%%%%%%
%
\section{Baryon asymmetry}
\label{sec:Baryon_asymmetry}
%
%%%%%%%%%%%%%%%%%%%%%%%%%%%%%%%%%%%%%%%%%%%%%%%%%%%%%%%%%%%%%%%%%%%%%%%%%%%%%%%%%%%%%%%%%%%%%%%%%%%
%%%%%%%%%%%%%%%%%%%%%%%%%%%%%%%%%%%%%%%%%%%%%%%%%%%%%%%%%%%%%%%%%%%%%%%%%%%%%%%%%%%%%%%%%%%%%%%%%%%

The goal of this paper is to compute the baryon number produced locally at a space-time position where wall has a velocity $v_w$. This assumption was integrated in our equations from the beginning, so all our chemical potentials are functions of $v_w$. The fundamental quantity of interest for this process is the left-chiral baryon chemical potential: $\xi_{B_\L} = \sum_{q} \xi_{qL}$. Given this seed asymmetry, the ensuing local baryon asymmetry follows from~\cite{Cline:2011mm}:
\begin{equation}
	\eta_B(v_w) = {405\,\hat \Gamma_{\rm sph}\over 4\pi^2 v_w\gamma_w g_*}\int d\hat z\, 
	\xi_{\!B_{\rm L}}f_{\rm sph}\,e^{-45\Gamma_{\rm sph}|z|/4v_w\!\gamma_w},
\label{eq:etab}
\end{equation}
where $\hat \Gamma_{\rm sph} = (18 \pm 3)\alpha_W^5 \approx 8\times 10^{-7}$~\cite{DOnofrio:2014rug} and the function $f_{\rm sph}(z) = {\rm min}(1,2.4\frac{\Gamma_{\rm sph}}{T}e^{-40h(z)/T})$ is designed to smoothly interpolate between the sphaleron rates in the broken and unbroken phases and $\hat z \equiv zT$. Finally $g_*$ is the number of degrees of freedom in the heat bath. We use the standard model value at very high temperatures: $g_*= 106.75$. 

We wish to emphasize that the true baryon asymmetry may not be represented well by $\eta_B(v_w)$ evaluated with a fixed $v_w$, because the for example the shock reheating may change the terminal wall velocities in an inhomogeneous way~\cite{Hindmarsh:2017gnf,Cutting:2019zws,Cutting:2020nla}. Because the baryon number can depend quite strongly on $v_w$, it would be desirable to compute the total baryon number as a weighted integral
\begin{equation}
\eta_B = \int_0^1 {\rm d}v_w P(v_w) \eta_B(v_w),
\label{eq:Bint_in_vw}
\end{equation}
where $P(v_w){\rm d}v_w$ is the fraction of the spatial volume swept by a phase transition wall moving with the velocity $v_w$, such that $\int_0^1{\rm d}v_w P(v_w)=1$. It is in fact quite possible that uncertainty about wall dynamics, formally quantified by $P(v_w)$ in~\cref{eq:Bint_in_vw}, will eventually dominate the error budget in the BAU prediction. 

Apart from the uncertainty in the total baryon number, the macroscopic spatial dependence of the wall velocity $v_w(x)$ could lead to a spatially varying baryon density $\eta_B(x)$, where the amplitude of the spatial variation could be quite large. The size of the inhomogeneities would be restricted to a fraction of the causal horizon during the electroweak phase transition however. Determining either $P(v_w)$ or $\eta_B(x)$ is beyond the scope of this paper, but they could be easily measured from a detailed simulation of the bubble growth and coalescence.

%%%%%%%%%%%%%%%%%%%%%%%%%%%%%%%%%%%%%%%%%%%%%%%%%%%%%%%%%%%%%%%%%%%%%%%%%%%%%%%%%%%%%%%%%%%%%%%%%%%
%%%%%%%%%%%%%%%%%%%%%%%%%%%%%%%%%%%%%%%%%%%%%%%%%%%%%%%%%%%%%%%%%%%%%%%%%%%%%%%%%%%%%%%%%%%%%%%%%%%
%
\section{The benchmark model}
\label{sec:bencmark-model}
%
%%%%%%%%%%%%%%%%%%%%%%%%%%%%%%%%%%%%%%%%%%%%%%%%%%%%%%%%%%%%%%%%%%%%%%%%%%%%%%%%%%%%%%%%%%%%%%%%%%%
%%%%%%%%%%%%%%%%%%%%%%%%%%%%%%%%%%%%%%%%%%%%%%%%%%%%%%%%%%%%%%%%%%%%%%%%%%%%%%%%%%%%%%%%%%%%%%%%%%%

To study how the baryon asymmetry depends on the number of moments and the truncation and factorization assumptions, we need to define a specific model. We choose the model defined in~\cite{Cline:2020jre}, which includes a heavy top quark, a massless left-chiral bottom quark and a massless higgs boson, coupled via gauge and Yukawa interactions.  The source for the CP-violation comes from the complex mass of the top-quarks, which is assumed to have the usual coupling to the higgs field $h$ and a dimension-5 operator $i(s/\Lambda) \bar Q_3 H t_\R$, where $s$ is another scalar field. When both fields vary across the wall, the top quark gets an effective spatially varying complex mass term:
\be
  m_t(z) = y_t h(z) \left( 1 + i {s(z)\over \Lambda}\right),
\label{mteq}
\ee
which implies
\begin{equation}
	|m_t(z)| = y_t h(z)\sqrt{1 + s^2(z)/\Lambda^2} \qquad {\rm and} \qquad
	\theta(z) = \tan^{-1}{s(z)\over\Lambda}\,.
\end{equation}
We will concentrate on the CP-odd sector and take a phenomenological approach, where $v_w$ is treated as a free parameter and the scalar fields $h(z)$, $s(z)$ are modeled as
\begin{equation}
	h(z) = {v_n\over 2}\left(1 - \tanh {z\over L_w}\right) \qquad {\rm and} \qquad
	s(z) = {w_n\over 2}\left(1 + \tanh {z\over L_s}\right).
\label{wall-profile}
\end{equation}
For the parameters defining these ansäze, we use
\begin{equation}
 v_n = \sfrac{1}{2} w_n = T_n, \qquad  
 \Lambda = 1\,{\rm TeV}, \qquad
  L_w = L_s = {5\over T_n}.
\label{eq:fiducial_model}
\end{equation}
For the nucleation temperature, we use $T_n = 100$ GeV. Given the $z$-dependent masses, one can compute the ensuing semiclassical force terms $\mathcal{S}_{t,b}$ from equation~\cref{eq:sterms}. What remains, is to define the precise collision equation network for the problem.

%%%%%%%%%%%%%%%%%%%%%%%%%%%%%%%%%%%%%%%%%%%%%%%%%%%%%%%%%%%%%%%%%%%%%%%%%%%%%%%%%%%%%%%%%%%%%%%%%%%
%
\subsection{Collision equation network}
\label{eq:coll_eq_network}
%
%%%%%%%%%%%%%%%%%%%%%%%%%%%%%%%%%%%%%%%%%%%%%%%%%%%%%%%%%%%%%%%%%%%%%%%%%%%%%%%%%%%%%%%%%%%%%%%%%%%

We will closely follow the treatment of~\cite{Cline:2020jre}, including four species in our equation network: the left and right helicity tops, left helicity bottom and a representative for the higgs fields, which are all assumed to be degenerate. Under these assumptions we get the following network of collision terms for the four species involved:
\begin{align}
\hat{\cal C}_{t_\pm}  &=  K^t_\ell {\hat C}_{q_\pm} 
  - u_\ell^{t_\pm} \kappa_{q_\pm} \hat\Gamma^{t_\pm}_{\rm tot} \mp \hat\Gamma_{\rm SS}[\xi]
\nn\\                 
\hat{\cal C}_{b_-}  &=  K^b_\ell {\hat C}_{b_-}  
               - u_\ell^{b_-} \kappa_{b_-} \hat\Gamma^{b_-}_{\rm tot} + \,\hat\Gamma_{\rm SS}[\xi]
\nn\\                 
\hat{\cal C}_h \phantom{l} &= K^h_\ell {\hat C}^{\rm Y}_{h} 
                  - u_\ell^h \kappa_h {\hat\Gamma}^h_{\rm tot},
\label{eq:collision_terms2a}
\end{align}
where $\kappa_i \equiv -1/\langle f_{0wi} \rangle = -N_1/n_i = |N_1|/n_i$ and the $K^a_\ell$-functions are defined in~\cref{{eq:K-function}}. The collision terms $\hat{\cal C}_{a}$ associated with the chemical potentials are given by
\def\tsum{\textstyle \sum}
\begin{align}
{\hat C}_\tL  &= \Gamma^t_y \,(\xi_\tL \!\!- \xi_\tR \! + \xi_h )      
                + \hat\Gamma_{\rm \sss W} (\xi_\tL \!\!- \!\xi_\bL)
                + \hat\Gamma^t_{\rm hf} (\xi_\tL \!\!-\! \xi_\tR) 
\nn\\                 
{\hat C}_\bL  &= \Gamma^t_y \,(\xi_\bL \!\!- \xi_\tR \! + \xi_{h} )      
              + \hat\Gamma_{\rm \sss W} (\xi_\bL \!\! - \!\xi_\tL) 
\nn\\                 
{\hat C}_\tR  &= \Gamma^t_y \,(2\xi_\tR \!\!- \xi_\tL \! - \xi_\bL \! - 2\xi_h)      
                + \hat\Gamma^t_{\rm hf} (\xi_\tR \!\!-\! \xi_\tL) 
\nn\\
{\hat C}_h \phantom{i} &= 3 \hat\Gamma^t_y \, (\xi_h  - \xi_\tR \! + \xi_\tL) 
                 + \hat\Gamma_{\rm \sss W} (\xi_h  \! - \xi_{h^+}) + \hat\Gamma_{h} \xi_{h}.
\label{eq:collision_terms2c}
\end{align}
The various rate coefficients appearing in~\cref{eq:collision_terms2a,eq:collision_terms2c} will be defined shortly below.

%%%%%%%%%%%%%%%%%%%%%%%%%%%%%%%%%%%%%%%%%%%%%%%%%%%%%%%%%%%%%%%%%%%%%%%%%%%%%%%%%%%%%%%%%%%%%%%%%%%
\paragraph{Strong sphaleron rate and the seed asymmetry}
%%%%%%%%%%%%%%%%%%%%%%%%%%%%%%%%%%%%%%%%%%%%%%%%%%%%%%%%%%%%%%%%%%%%%%%%%%%%%%%%%%%%%%%%%%%%%%%%%%%

The strong sphaleron rate defined in~\cref{eq:SSR1} contains chemical potentials for all quarks. In particular for the light quarks from two first generations this is the only relevant interaction rate. This implies that all light quarks have the same chemical potential obeying $\xi_{q_\L} = -\xi_{q_\R} \equiv \xi_q$. In our setup this condition applies also to the right helicity bottom quark $b_\tR$. Setting $B = \sum_q (n_q-\bar n_q) = 0$ then gives a constraint
\be
\mu_{q} = D_0^t(\mu_{\tL} \!+\!\mu_{\tR}) + \mu_{\bL},
\label{eq:lightquarks}
\ee
where $D_0 = \langle f_{0w}^\prime \rangle$ is given by equation~\eqref{eq:D-and-Q} with $\ell = 0$. We set $D_0^b\equiv1$ in~\cref{eq:lightquarks}, because we assume $m_b=0$ throughout. Using~\eqref{eq:lightquarks} we can express the strong sphaleron rate in the benchmark model as:
\begin{equation}
\hat\Gamma_{\rm SS}[\xi] = \hat\Gamma_{\rm SS} \left[ \big(9D_0^t + 1\big) \xi_\tL 
+ \big(9D_0^t - 1\big) \xi_\tR  + 10\xi_\bL\right]. 
\label{eq:strongsph}
\end{equation}
Clearly this expression is dependent on the assumptions we made and indeed, a different result is obtained for example when one assumes that $b$-quark mass does not vanish and $b_+$ is taken to be a part of the reaction network.

As emphasized above, the essential physical quantity coming from solving the moment equations, is the left-handed baryon chemical potential: $\xi_{B_\L} = \sum_{q} \xi_{q_L}$, which drives the weak anomaly to create the baryon asymmetry. In the current setup this quantity can be written in terms of $\xi_{t_\pm}$ and $\xi_\bL$ as follows:
\begin{equation}
\xi_{B_\L} = \frac{1}{2}\big(1 + 4D_0^t) \xi_\tL + 2D_0^t\xi_\tR + \frac{5}{2}\xi_\bL.
\label{eq:xiBL_reduced}
\end{equation}
where we again used $D_0^b = 1$, assuming massless bottom quark.

Finally, we note that as no interaction in our equation network breaks the conservation of baryon number, we must have $\smash{B = \sum_q (n_q-n_{\bar q})=0}$ at all times. This implies a constraint $\smash{\hat C_\tL + \hat C_\tR + \hat C_\bL = 0}$, which is indeed satisfied by collision terms~\cref{eq:collision_terms2a}. Equation~\cref{eq:final_eq} is an easily adaptable system to any physical problem and for any number of moments. The only model dependent parts are related to the collision equation networks, the precise expressions for the reduced sphaleron rate~\cref{eq:strongsph}, and the final expression for the seed asymmetry~\cref{eq:xiBL_reduced}.

%%%%%%%%%%%%%%%%%%%%%%%%%%%%%%%%%%%%%%%%%%%%%%%%%%%%%%%%%%%%%%%%%%%%%%%%%%%%%%%%%%%%%%%%%%%%%%%%%%%
\paragraph{Benchmark rate coefficients}
%%%%%%%%%%%%%%%%%%%%%%%%%%%%%%%%%%%%%%%%%%%%%%%%%%%%%%%%%%%%%%%%%%%%%%%%%%%%%%%%%%%%%%%%%%%%%%%%%%%

We now define the various rate coefficients appearing in~\cref{eq:collision_terms2a,eq:collision_terms2c}. 
Computing and summing over all scattering processes as described in Section~\cref{sec:collision-terms} is an extensive task, which we will perform in~\cite{KimmoNiyati2}. Here we will instead use the simple estimates from the literature. For the strong sphaleron rate we use the value given in~\cite{Moore:1997im,Moore:2010jd}: $\hat \Gamma_{\rm SS} = 2.7\times 10^{-4}$. For the other interaction rates, only estimates exist, some decades old~\cite{Huet:1995sh}: $\hat\Gamma_y=4.2\times 10^{-3}$, $\hat \Gamma_m=x^2_t/63$ and $\Gamma_h=x^2_W/50$, where $x_t = m_t/T$ is the scaled top mass with $m_t$ given in~\eqref{mteq}, and $x_W \equiv gh(z)/2T$. Furthermore, for the total interaction rates we use~\cite{Fromme:2006cm}: $\hat \Gamma^i_{\rm tot} = -(D_{2,i}/D_{1,i})v_w/\hat D_i$ where $D_{\ell,i}$ are the moment functions defined in~\cref{eq:D-and-Q} and in~\cref{eq:moment_functions_appendix}, and the dimensionless quark and higgs diffusion constants are $\hat D_q=6$ and $\hat D_h = 20$, respectively. These approximations suffice here, because we are mainly interested in the convergence of the moment expansion and its truncation and factorization rule dependencies, and this lets us compare our results to the existing literature. To this end we also set $\kappa_a$-factors to unity in elastic rates. However, we still differ slightly from~\cite{Cline:2020jre} by our more consistent use of the MB-approximation.

%%%%%%%%%%%%%%%%%%%%%%%%%%%%%%%%%%%%%%%%%%%%%%%%%%%%%%%%%%%%%%%%%%%%%%%%%%%%%%%%%%%%%%%%%%%%%%%%%%%
%%%%%%%%%%%%%%%%%%%%%%%%%%%%%%%%%%%%%%%%%%%%%%%%%%%%%%%%%%%%%%%%%%%%%%%%%%%%%%%%%%%%%%%%%%%%%%%%%%%
%
\section{Numerical results}
\label{sec:numerical_results}
%
%%%%%%%%%%%%%%%%%%%%%%%%%%%%%%%%%%%%%%%%%%%%%%%%%%%%%%%%%%%%%%%%%%%%%%%%%%%%%%%%%%%%%%%%%%%%%%%%%%%
%%%%%%%%%%%%%%%%%%%%%%%%%%%%%%%%%%%%%%%%%%%%%%%%%%%%%%%%%%%%%%%%%%%%%%%%%%%%%%%%%%%%%%%%%%%%%%%%%%%

\begin{figure}
\centering
\includegraphics[width=0.93\textwidth]{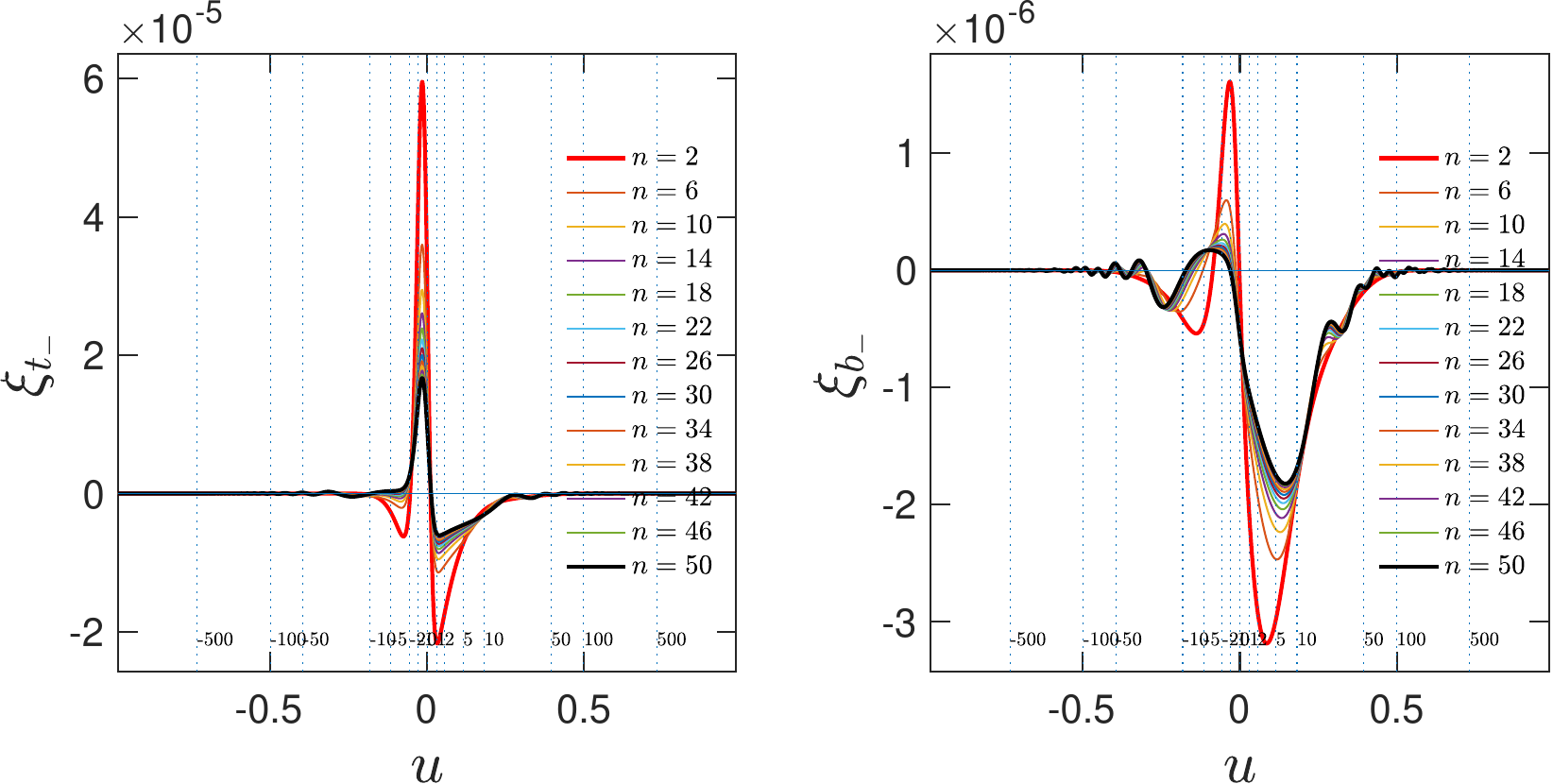}\hskip 0.5truecm
\caption{Chemical potentials for $\tL$ (left panel) and for $\bL$ (right panel) for the benchmark model with $v_w = 0.1$, obtained from runs with different number of moments as shown in the legends.}
\label{fig:chemical_potentials}
\end{figure}

In this section we present our numerical results. We study the dependence of baryon asymmetry and chemical potentials on the number of moments and the wall velocity, as well as on the factorization and truncation schemes defined in section~\cref{sec:moments}. We start by solving~\cref{eq:final_eq} for the benchmark model for $v_w = 0.1$ using the truncation $R = -v_w$ and the factorization rule~\cref{eq:barR} for $\bar R$. In figure~\cref{fig:chemical_potentials} we show the $t_-$ and $b_-$ chemical potentials as a function of the scaled variable%
%
% FOOTNOTE
%
\footnote{We have re-scaled the dimensionless spatial coordinate $zT$ into a new coordinate $u \in [-1,1]$. The re-scaling is designed to put more grid-points near the wall and progressively less points to far from the wall, where solutions are decaying exponentially. This allows us to get accurate results with much fewer grid points than in the case of linear $z$-grid. The precise form of the scaling function is not relevant however, and our numerical routines automatically distribute more points where needed.}  
%
% END FOOTNOTE
%
$u$ from calculations with different maximum number of moments: $n = 2,6,10,...,50$. The solid red curves correspond to a case with the lowest nontrivial number of moments $n=2$, and black solid curves to a case with the highest number of moments $n=50$. The number of moments for $n>2$ were chosen with the particular interval $\Delta n=4$ due to reasons explained in section~\cref{sec:moment_sequences}. The thin dotted vertical lines display the value of $zT$ in units of wall width $L_wT$.

%==================================================================================================
%==================================================================================================

%==================================================================================================
%==================================================================================================

\begin{figure}
\centering
\includegraphics[width=0.93\textwidth]{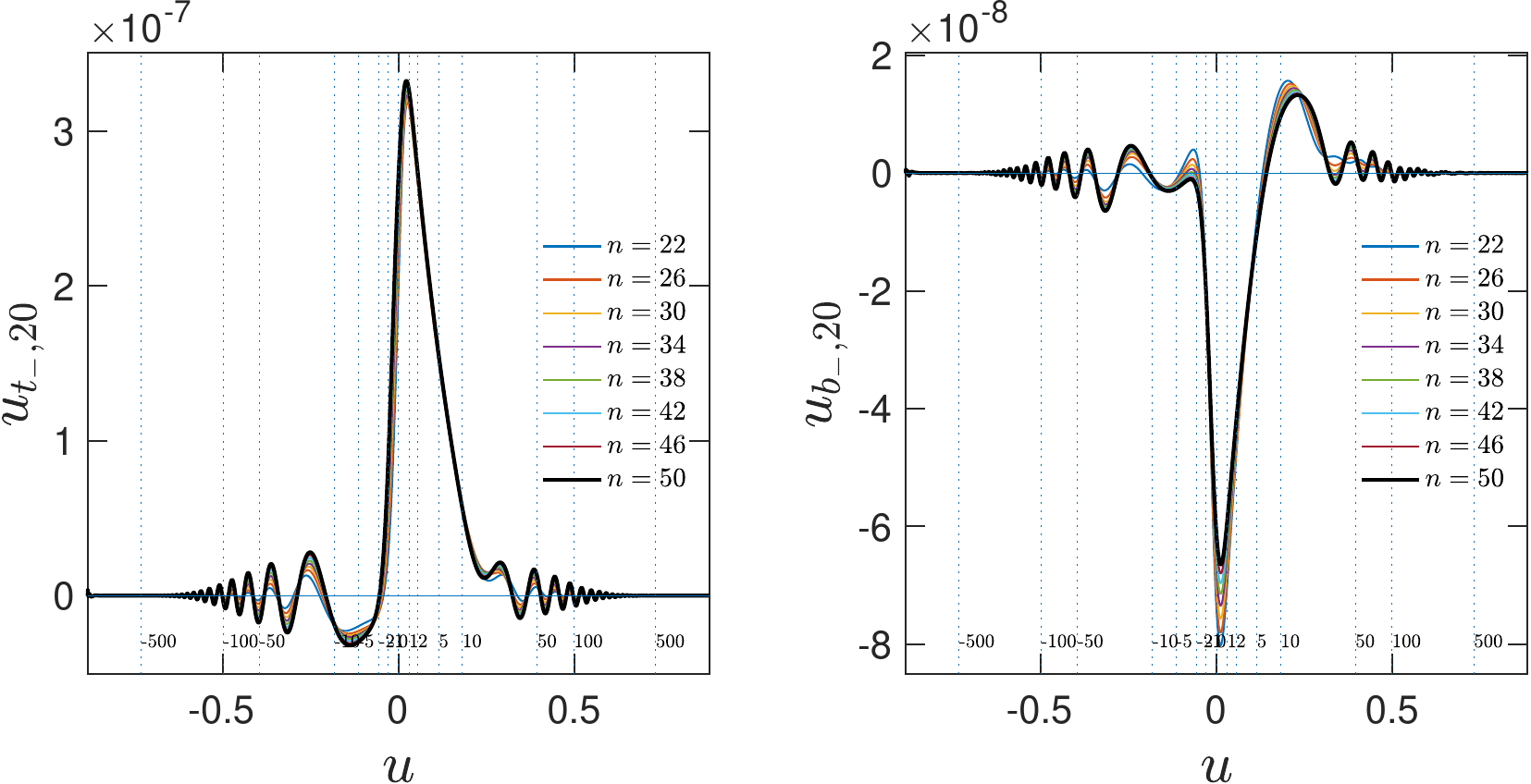}\hskip 0.5truecm
\caption{The 20'th moments $u_{q,20}$ for $\tL$ (left panel) and $\bL$ (right panel) for the same parameters as in figure~\cref{fig:chemical_potentials}.}
\label{fig:moment}
\end{figure}

Extending the evolution equations to include higher moments significantly changes the results for chemical potentials: the solutions tend to become less sharply varying near the wall, and they develop oscillatory behaviour at large distances from the wall. The clustering of higher moment curves shows that the solutions converge, however. Oscillations are more clearly visible in the bottom quark case, and in both cases they extend rather far in front of the wall, although this behaviour is masked in figure~\cref{fig:chemical_potentials} due to their very small amplitude. The oscillations appear more prominently in solutions for higher moments. In figure~\cref{fig:moment} we show for example the 20'th moment of the $\tL$ and $\bL$ distributions. This moment is of course available only in networks with $n\ge 20$, which explains the smaller number of curves in plots. Again the convergence of the solutions for higher moments is evident. 

%==================================================================================================
%==================================================================================================

%==================================================================================================
%==================================================================================================

%==================================================================================================
%==================================================================================================

\begin{figure}
\centering
\includegraphics[width=0.93\textwidth]{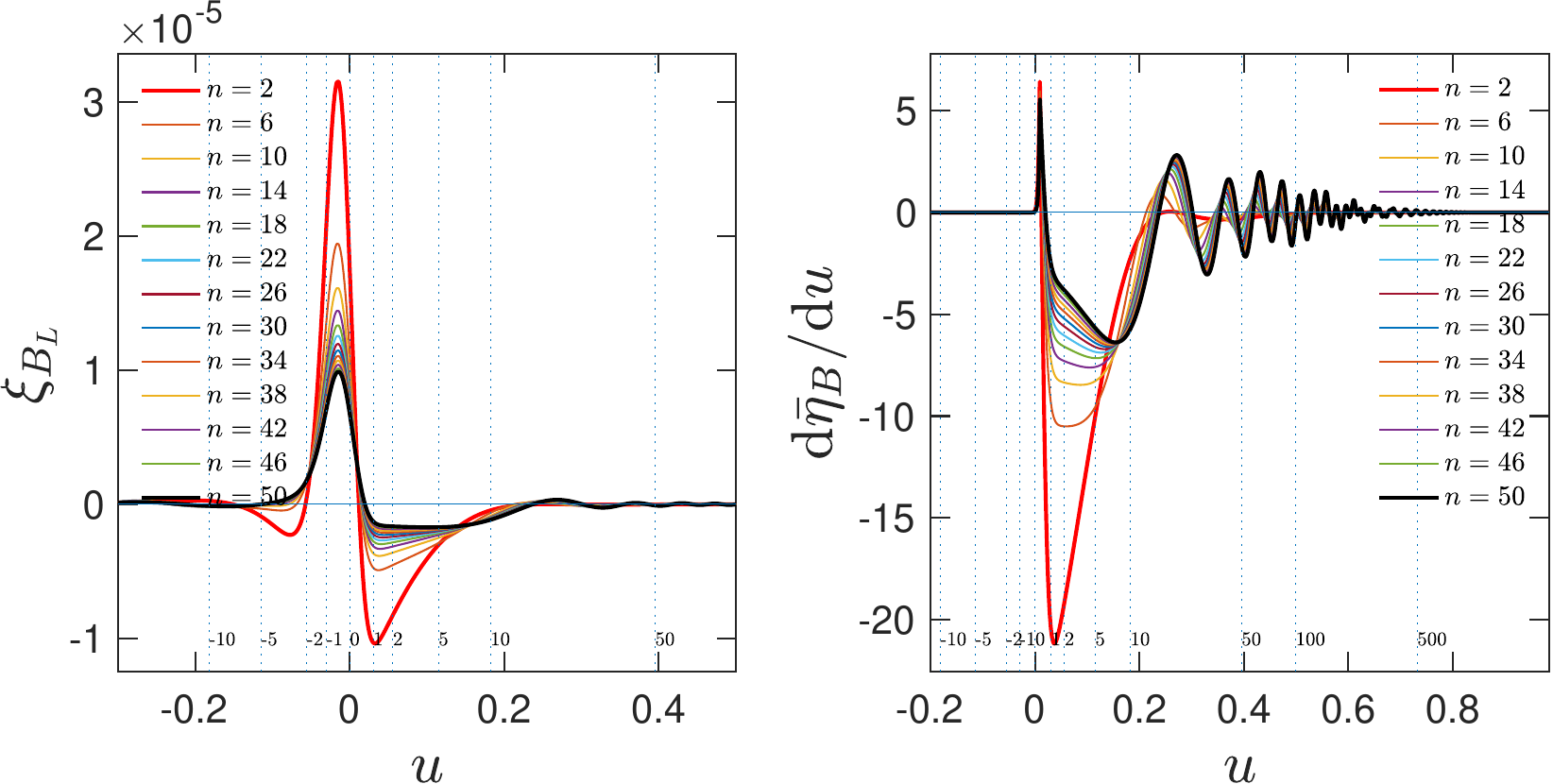}\hskip 0.5truecm
\caption{The seed asymmetry $\xi_{B_L}$ (left panel) and the differential ${\rm d}\bar\eta_B/{\rm d}u$, where $\bar \eta_{\rm B} \equiv \eta_{\rm B}/8.7\times 10^{-11}$ (right panel) for the same parameters as in figure~\cref{fig:chemical_potentials}.}
\label{fig:seed}
\end{figure}

%==================================================================================================
%==================================================================================================

Of course, the higher moment functions, or even the individual chemical potential distributions, are not of direct interest to us. The relevant quantity is the chemical potential $\xi_{B_{\rm L}}$ given in~\cref{eq:xiBL_reduced}, which biases the baryon number violating sphaleron processes. We show $\xi_{B_{\rm L}}$ in the left panel of figure~\cref{fig:seed} for our test case. This plot is not too dissimilar from the $\tL$ chemical potential shown in~\cref{fig:chemical_potentials}, except for slightly more flattening of the solution near wall and a more extended oscillatory reach away from the wall. However, oscillations are still somewhat masked by their quickly decaying amplitude. Their true extent becomes evident in the right plot in~\cref{fig:seed}, where we show the integrand ${\rm d}\eta_B/{\rm d}u$ of the expression~\cref{eq:etab}. Here the magnitude of the oscillations is amplified by their spatial extent encoded in the scaling function ${\rm d}(zT)/{\rm d}u$. We point out that the decay of the oscillatory tail is not due to the exponential suppression factor in ${\rm d}\eta_B/{\rm d}u$, but due to decay of $\xi_{B_{\rm L}}$ at large distances.

%==================================================================================================
%==================================================================================================

\begin{figure}
\centering
\includegraphics[width=0.93\textwidth]{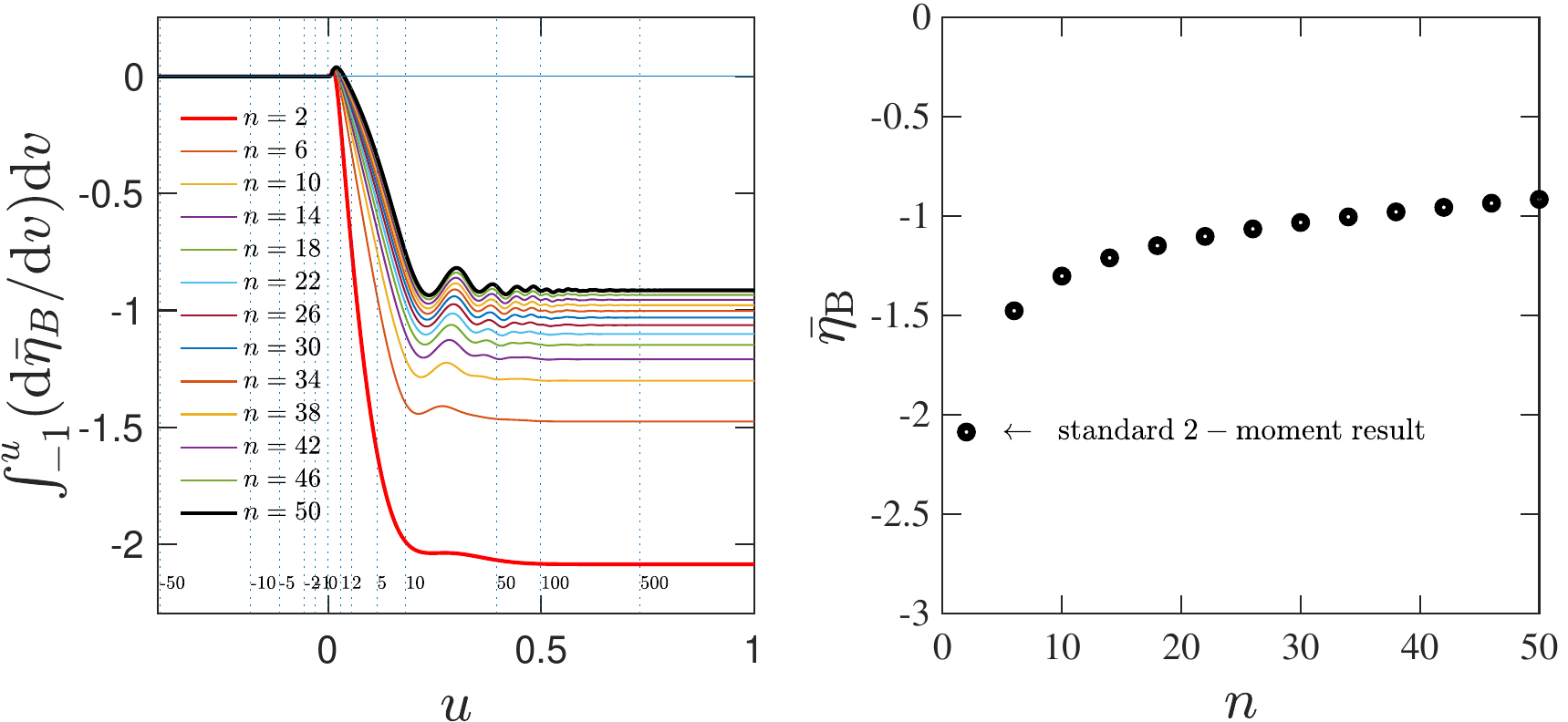}\hskip 0.5truecm
\caption{The cumulative baryon asymmetry integral in scaled variable $u$ (left panel) and the final integrated baryon asymmetry $\eta_B$ as a function of moments used in the calculation (right panel).}
\label{fig:cumulative_B}
\end{figure}

%==================================================================================================
%==================================================================================================

Even though oscillations show up quite prominently in the plot for ${\rm d}\eta_B/{\rm d}u$, they do not affect the final baryon asymmetry significantly. We show this by plotting the cumulative asymmetry $\smash{\int_{-1}^u ({\rm d}\eta_B/{\rm d}v}){\rm d}v$ as a function of $u$ in left panel of figure~\cref{fig:cumulative_B}. As we see, only the first few oscillations affect the cumulative asymmetry noticeably, and the contributions from adjacent oscillations tend to cancel out. While both the number of oscillations and their amplitude and extent tend to increase with $n$, the final asymmetry flattens out at around few hundred wall widths. The extent of the oscillations is still somewhat surprisingly large, given that the largest naive diffusion length in this problem is $\ell_{D_h} \sim D_h/v_w \approx 40L_w$. The reason for this phenomenon is probably the tendency of high energy modes to diffuse over longer distances than the low energy modes: the higher moments probe the higher energy spectrum of the perturbation $\delta f(p)$, and this shows up in more slowly decaying eigensolutions to the asymptotic equations, as we shall see explicitly in section~\cref{sec:eigenmode_analysis} below. At any rate, all evolution is taking place well below the maximum spatial distance, which in this calculation was set to $|z_{\rm max}| = 10^4 L_w$.

In the right panel of figure~\cref{fig:cumulative_B} we finally show the total baryon asymmetry $\eta_B$ as a function of moments used in the calculation. The results depend smoothly on $n$ and the asymmetry shows a converging tendency, but does not yet reach a quite clear asymptotic value with $n=50$. This gradual convergence is not due to the oscillations, but due to slow convergence of the solution $\xi_{B_{\rm L}}$ as a function of $n$ in the region $0 \lsim zT \lsim 5$, as can be seen from the evolution of the cumulative asymmetry shown in the left panel. In section~\cref{sec:truncation_dependence} we shall see that for small wall velocities the results depend also on the truncation choice, and in section~\cref{sec:moment_sequences} we find additional scatter due to different choice of moment sequences apart from the sequence $n=2,6,10,...$ used here. But let us first study the origin of the oscillations.

 %==================================================================================================
%==================================================================================================

\begin{figure}
\centering
\includegraphics[width=0.3\textwidth]{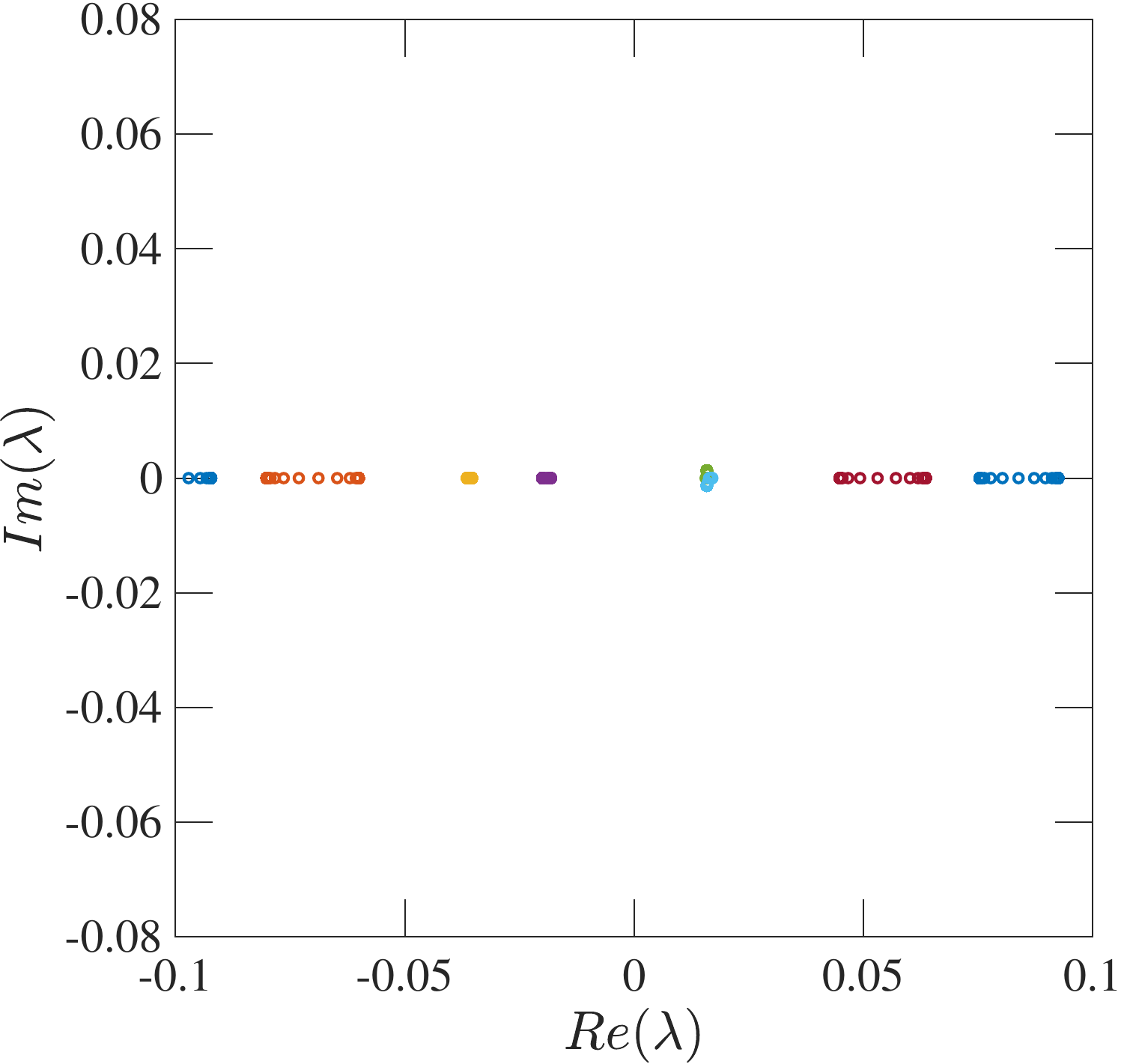}\hskip 0.3truecm
\includegraphics[width=0.3\textwidth]{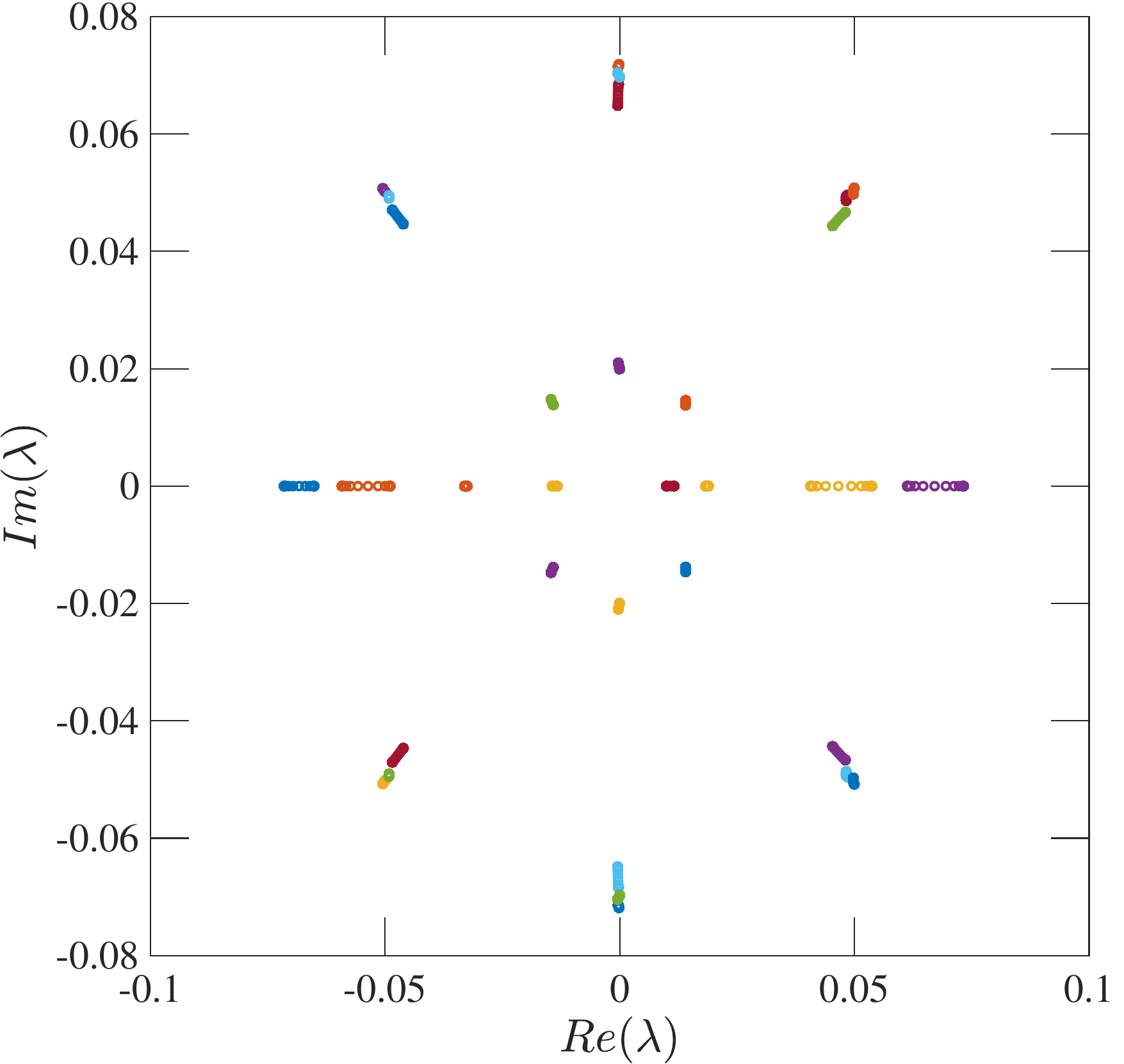}\hskip 0.3truecm
\includegraphics[width=0.3\textwidth]{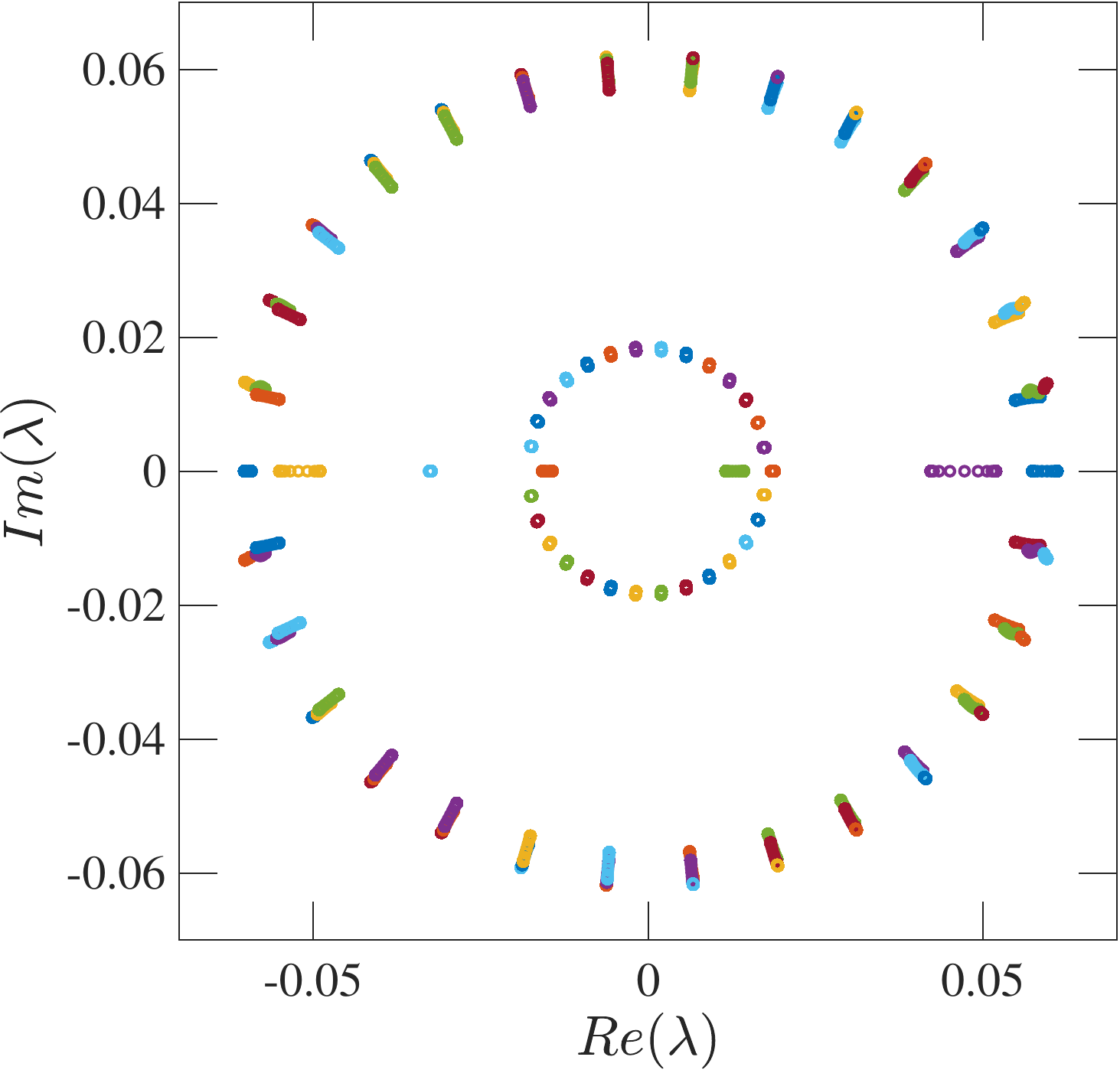}\hskip 0.3truecm
\caption{The eigenvalues of the asymptotic matrix $\hat {\cal X}$ in the differential equation  $\mathcal{W}^\prime= \hat{\mathcal{X}}\mathcal{W}$ in the case $n=2$ (left panel), $n=8$ (middle panel) and and in the case $n=30$ (right panel).}
\label{fig:eigen_values}
\end{figure}

%==================================================================================================
%==================================================================================================

%%%%%%%%%%%%%%%%%%%%%%%%%%%%%%%%%%%%%%%%%%%%%%%%%%%%%%%%%%%%%%%%%%%%%%%%%%%%%%%%%%%%%%%%%%%%%%%%%%%
%
\subsection{Oscillating eigenmodes}
\label{sec:eigenmode_analysis}
%
%%%%%%%%%%%%%%%%%%%%%%%%%%%%%%%%%%%%%%%%%%%%%%%%%%%%%%%%%%%%%%%%%%%%%%%%%%%%%%%%%%%%%%%%%%%%%%%%%%%

To understand the reason for the spatial oscillations in the solutions, we make an eigenvalue analysis of the asymptotic differential operator associated with the fluid equations. Indeed, equation~\cref{eq:deqs} can be written asymptotically as $\mathcal{W}^\prime=\hat{\mathcal{A}}^{-1}\hat\Gamma \mathcal{W} \equiv \hat{\mathcal{X}}\mathcal{W}$. The $Nn\!\times\! Nn$-matrix $\hat{\mathcal{X}}$ is real but not symmetric, so it has $Nn$ eigenvalues $\lambda_i$ which are either real or appear in complex conjugate pairs. In the asymptotic region $|z|\rightarrow \infty$ the matrix $\hat{\mathcal{X}}$ becomes a constant, after which the eigenfunctions of $\hat{\mathcal{X}}$ behave as $w_i \sim w\exp(\lambda_i z)$. The eigenvectors corresponding to $\mathcal{R}(\lambda_i)>0$ are then exponentially growing when $z\rightarrow \infty$ and must be set to zero in the right boundary. Similarly, those corresponding to $\mathcal{R}(\lambda_i)<0$ must be set to zero in the left boundary. However, as the eigenvalues are in general complex, the corresponding solutions show a characteristic pattern of damped oscillators. 

It turns out that as the number of moments increases, more and more slowly damped but rapidly oscillating solutions emerge. In the left panel of figure~\cref{fig:eigen_values}, we show the eigenvalues of the matrix $\hat{\mathcal{X}}$ in the case $n=2$. Each differently colored set of points shows the evolution of a given eigenvalue across the transition wall in our benchmark scenario. Clearly all eigenvalues are real in this case%
%
% FOOTNOTE BEGINS
%
\footnote{Except at one singular point where the two eigenvalues $\lambda_i \approx 0.0135$ cross and get complex parts as required by the symmetry of the matrix $\hat{\mathcal{X}}$.}. 
%
% FOOTNOTE ENDS
%
In the rightmost panel, showing the similar evolution of eigenvalues in the case $n=30$, we see a very large number of complex eigenvalues in a symmetric pattern. The distance between eigenvalues in the concentric circles decreases as $n$ increases, giving rise to rapidly oscillating and slowly damped eigenfunctions. As these solutions couple to or contain fractions of lower moments and chemical potentials, they can cause the observed oscillations in the seed asymmetry. In the middle panel, which corresponds to $n=8$, we see an even more problematic set of eigenvalues with $\Re(\lambda) \approx 0$. We will return to this case in more detail in section~\cref{sec:moment_sequences} below.

%==================================================================================================
%==================================================================================================

\begin{figure}
\centering
\includegraphics[width=0.40\textwidth]{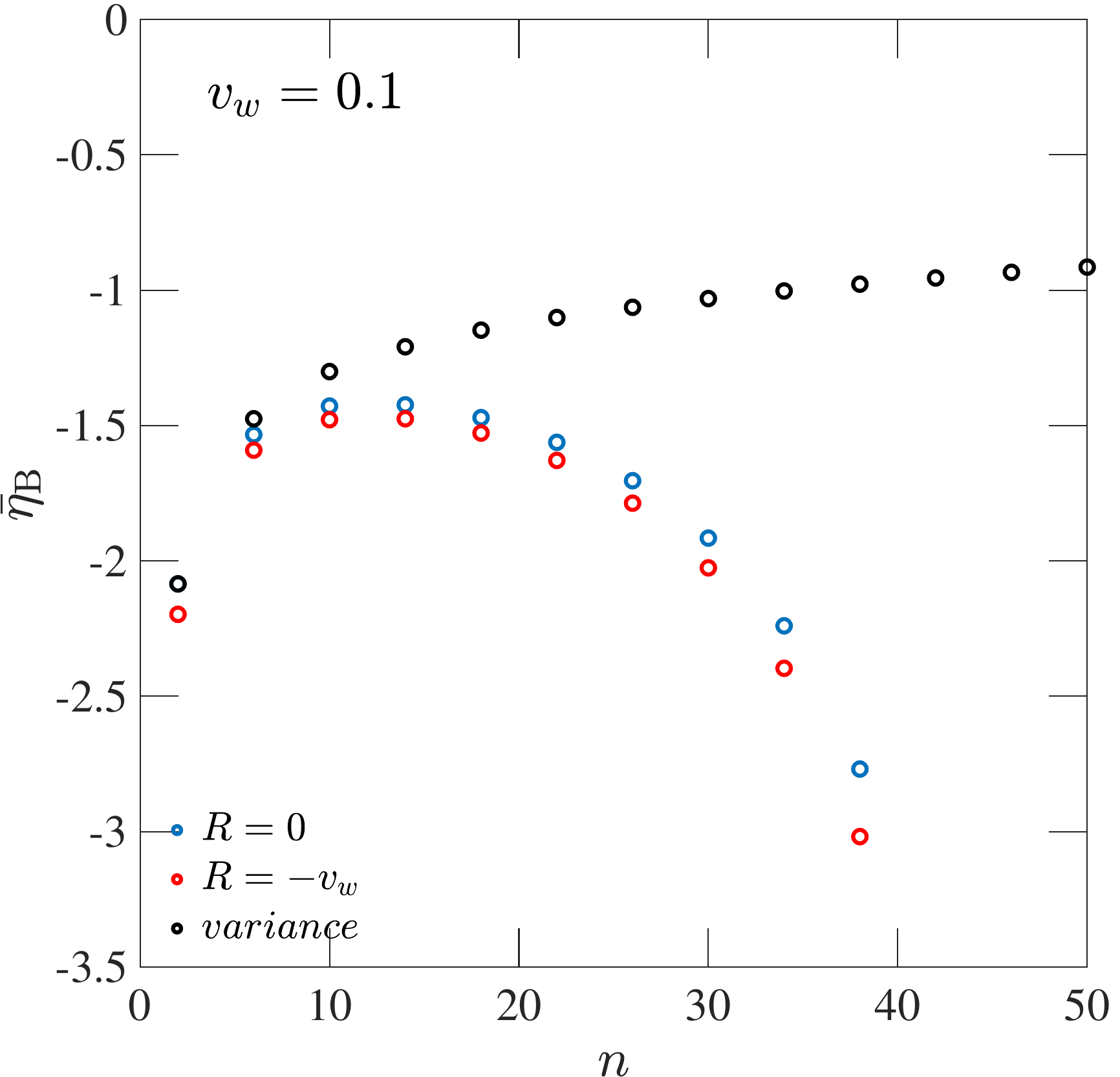}\hskip 0.5truecm
\includegraphics[width=0.40\textwidth]{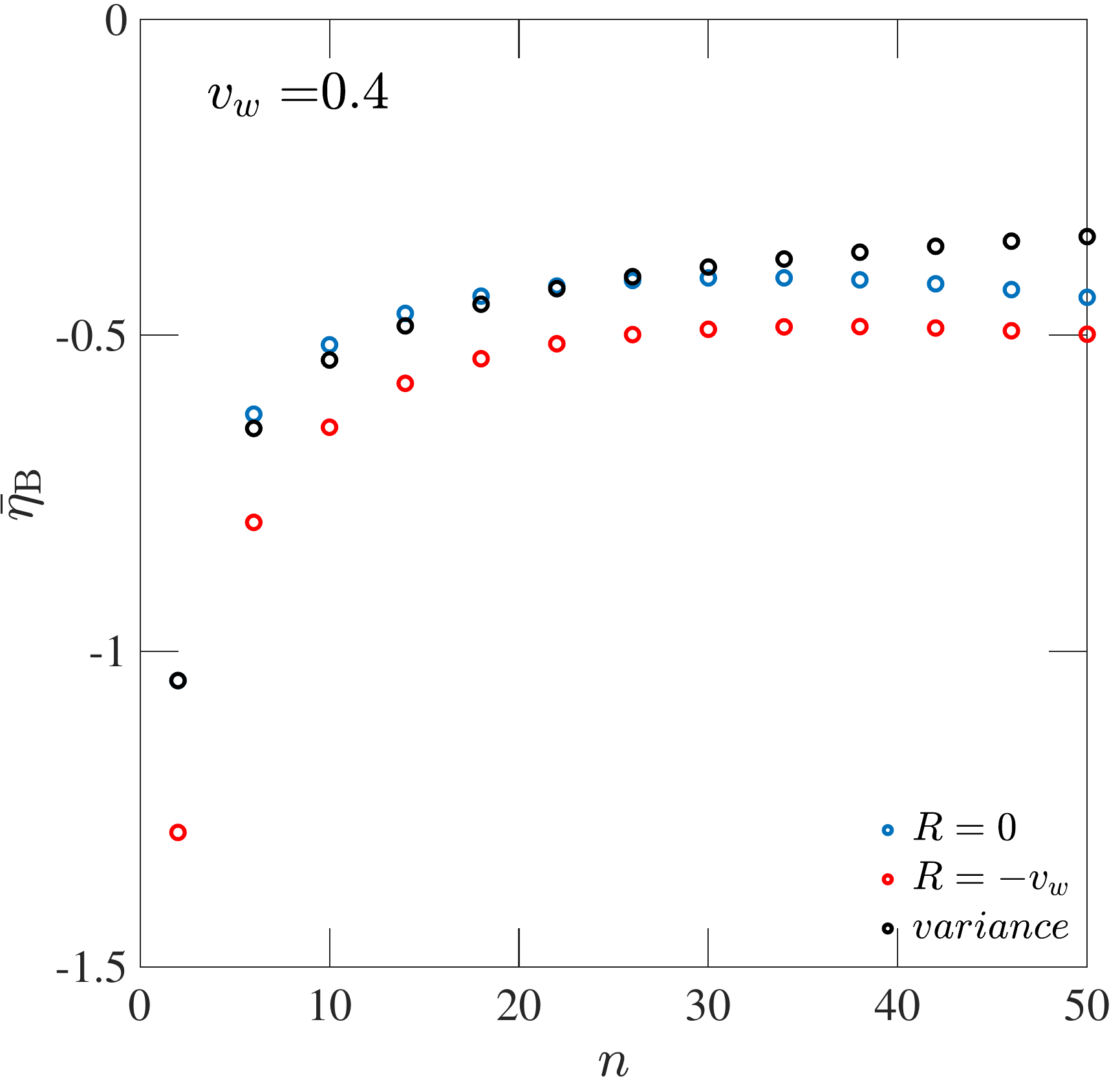}\hskip 0.5truecm
\includegraphics[width=0.40\textwidth]{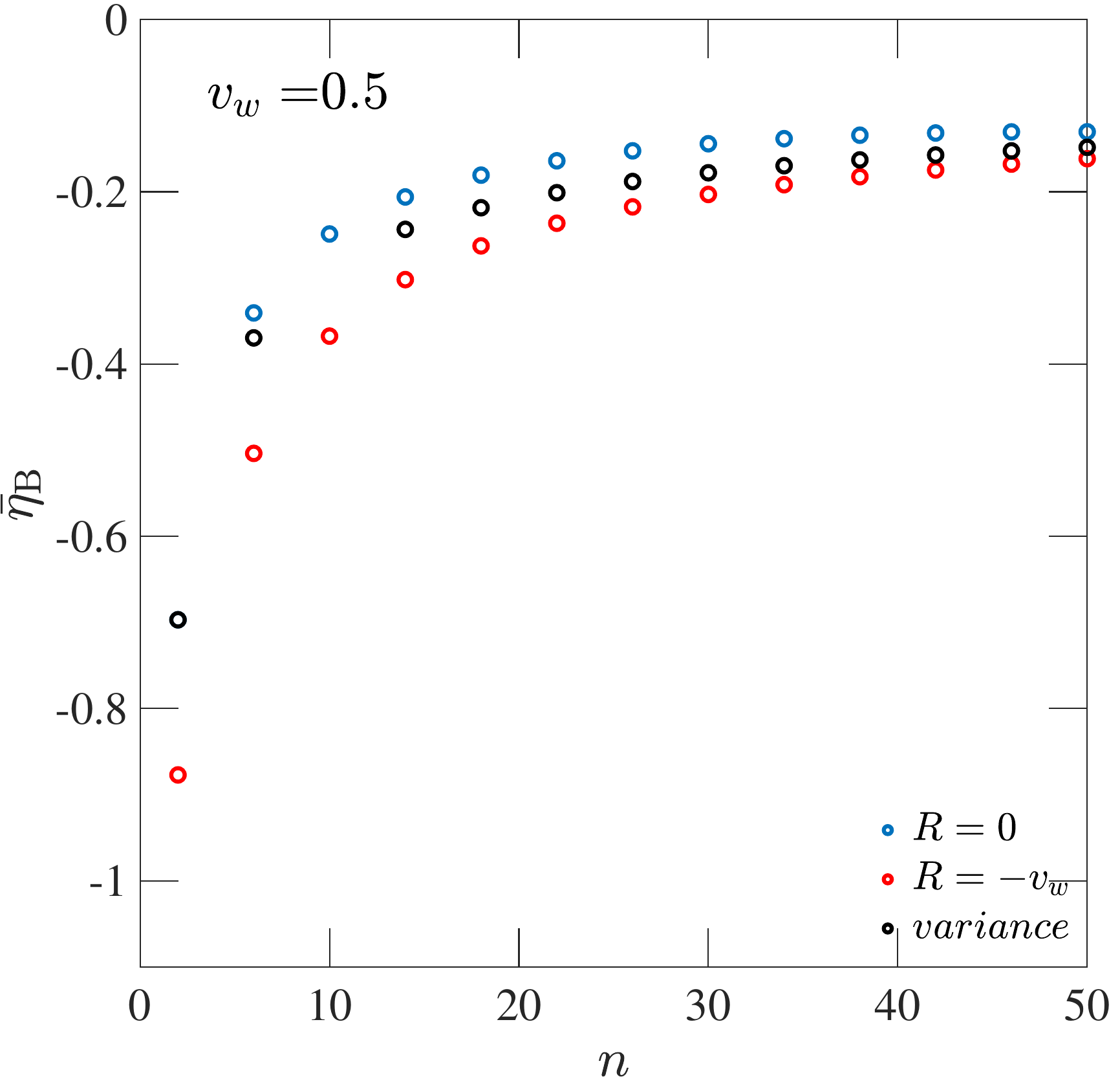}\hskip 0.5truecm
\includegraphics[width=0.40\textwidth]{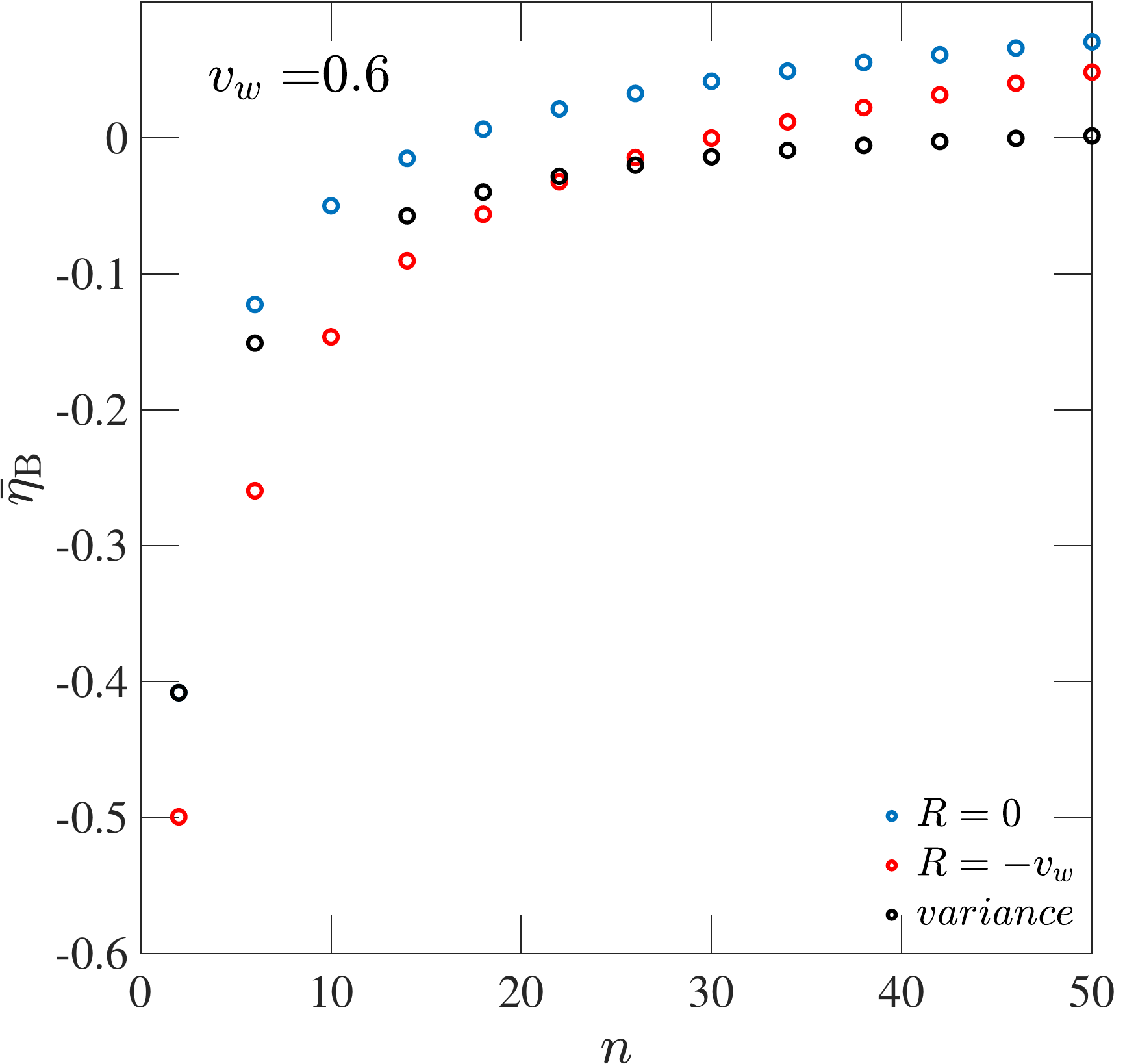}\hskip 0.5truecm
\caption{We show the predicted baryon asymmetry as a function of the number of moments for different wall velocities $v_w=0.1$, 0.4, 0.5 and 0.6 with different truncation choices.  In each panel black ellipses correspond to the variance truncation scheme~\cref{eq:R}, while the blue and red ellipses correspond to the constant truncation scheme~\cref{eq:trunction-condition} with $R=0$ and $R=-v_w$, respectively.}
\label{fig:vw-and-truncation-dependence}
\end{figure}

%==================================================================================================
%==================================================================================================

%%%%%%%%%%%%%%%%%%%%%%%%%%%%%%%%%%%%%%%%%%%%%%%%%%%%%%%%%%%%%%%%%%%%%%%%%%%%%%%%%%%%%%%%%%%%%%%%%%%
%
\subsection{Truncation and factorization dependence}
\label{sec:truncation_dependence}
%
%%%%%%%%%%%%%%%%%%%%%%%%%%%%%%%%%%%%%%%%%%%%%%%%%%%%%%%%%%%%%%%%%%%%%%%%%%%%%%%%%%%%%%%%%%%%%%%%%%%

We have also studied how the dependence on the factorization and truncation choices. We found that the factorization assumption, relating to the choice of the $\bar R$-parameter in equation~\cref{eq:genericmomeq}, has very little effect on the baryon asymmetry, and we do not display specific results here. Solutions are much more sensitive on the truncation choice however, in particular for the small wall velocities.  In figure~\cref{fig:vw-and-truncation-dependence} we show how the final baryon asymmetry depends on the number of moments for a number of different wall velocities and truncation choices. In the upper left panel we study again the case $v_w = 0.1$, showing the old result with variance truncation in black ellipses. In addition we display the result obtained with constant truncation schemes $R=0$, corresponding to setting $u_{n+1}'=0$ (blue ellipses), and $R=-v_w$, corresponding to setting $u_{n+1}'=-v_w u_n'$ (red ellipses). For relatively few moments the results agree, but after $n=12-16$ moments the constant truncation solutions start predicting larger and larger magnitude BAU as $n$ increases. The moment expansion thus fails to give an accurate prediction for the BAU in this truncation scheme.

For larger wall velocities, the situation improves. First, with $v_w=0.4$ (upper right panel) and $v_w=0.5$ (lower left panel) the downward turn as a function of $n$ in the constant truncation solutions stops, and solutions with different truncation schemes tend to get closer to each other. In the case $v_w=0.5$ the truncation dependence has all but vanished, and the solutions have developed a wide plateau. However, in the case $v_w=0.6$ (lower right panel) the spread with different truncation schemes has started to increase again. Anyway, the spread in the predicted baryon asymmetry is relatively small, of order $\Delta \bar\eta_{\rm B} \sim 0.1$, over a range in (a sufficiently large) $n$ and over different truncation methods. 

%==================================================================================================
%==================================================================================================

\begin{figure}
\centering
\includegraphics[width=0.40\textwidth]{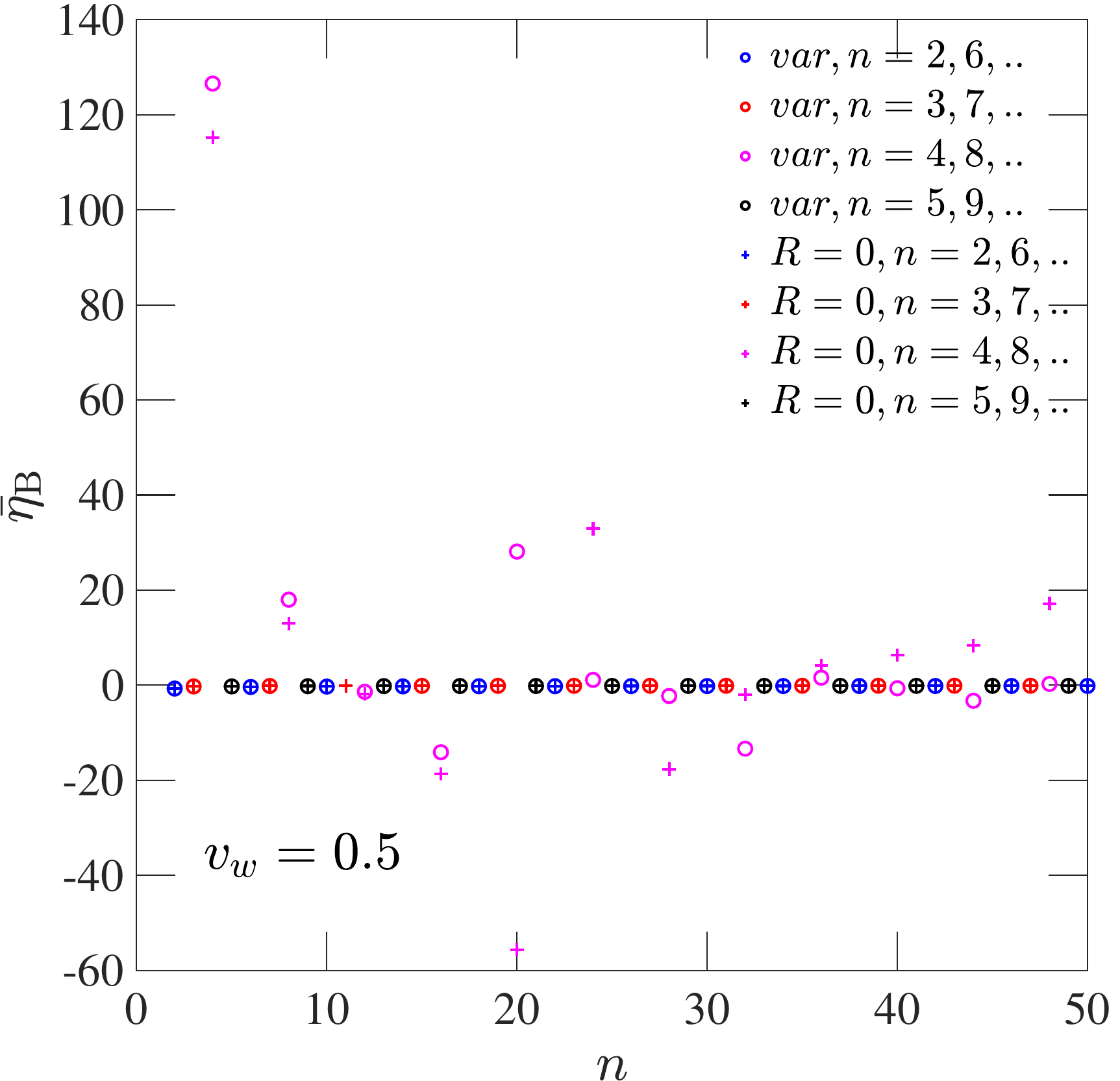}\hskip 0.5truecm
\includegraphics[width=0.40\textwidth]{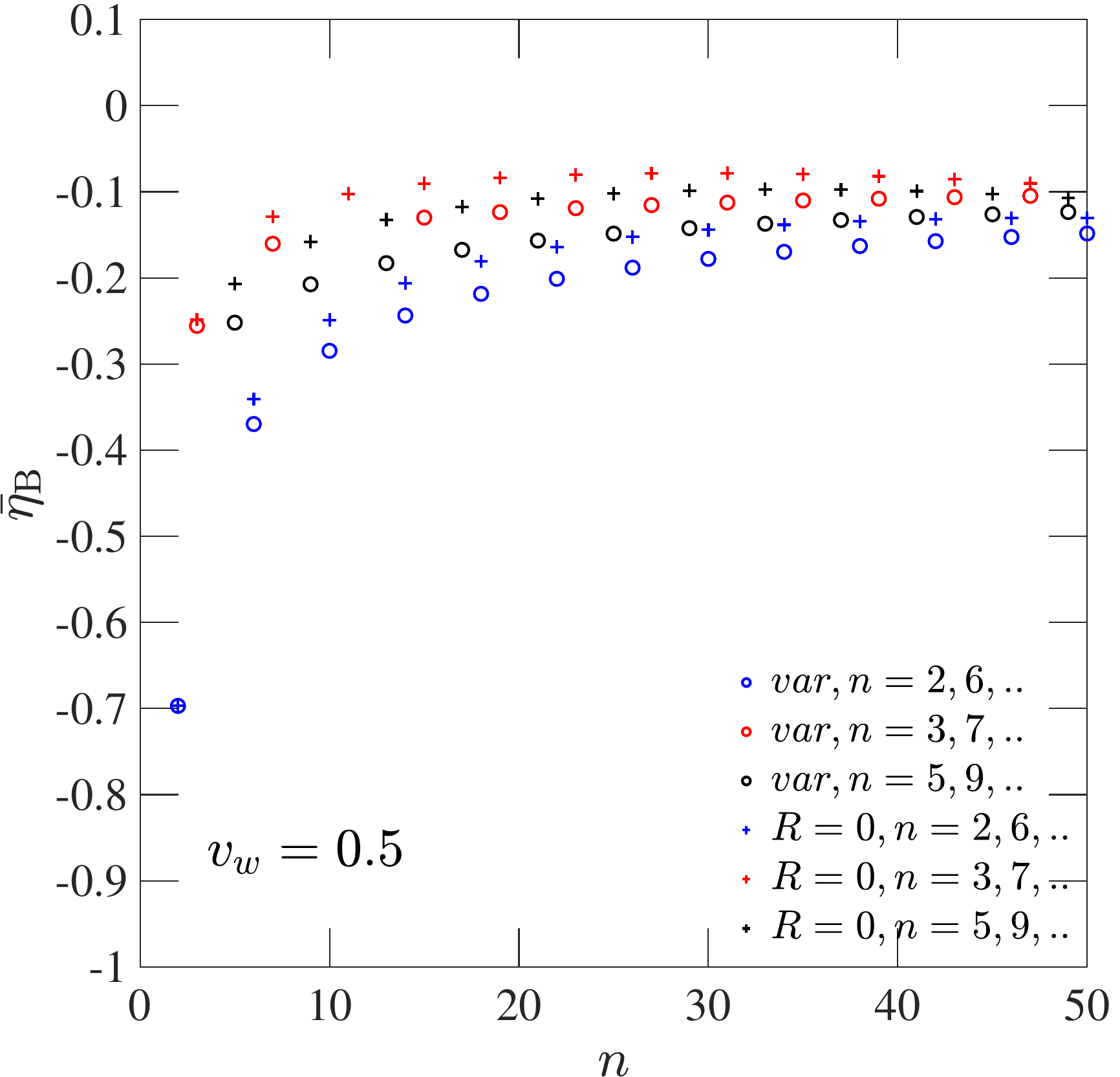}\hskip 0.5truecm
\includegraphics[width=0.40\textwidth]{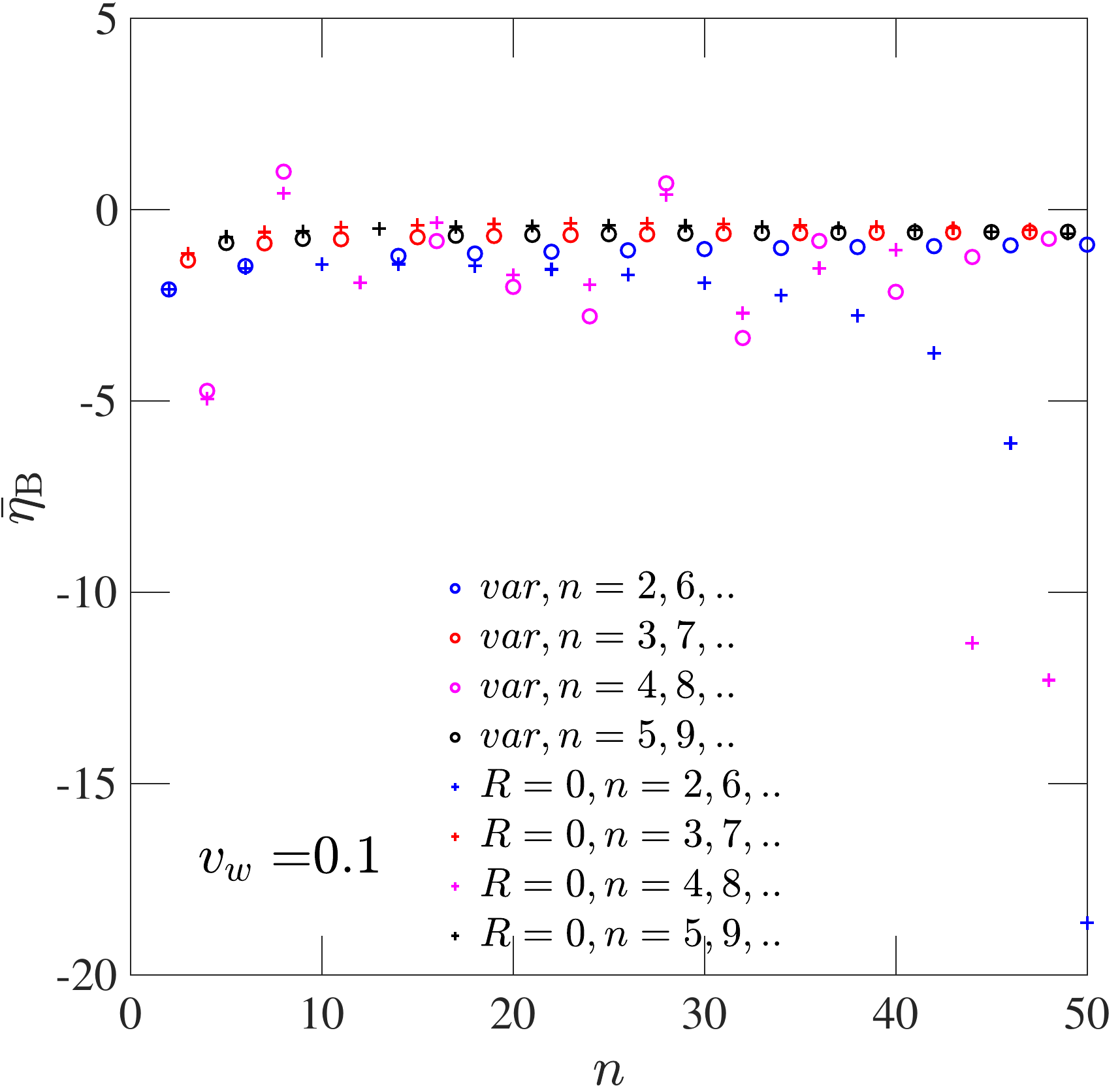}\hskip 0.5truecm
\includegraphics[width=0.40\textwidth]{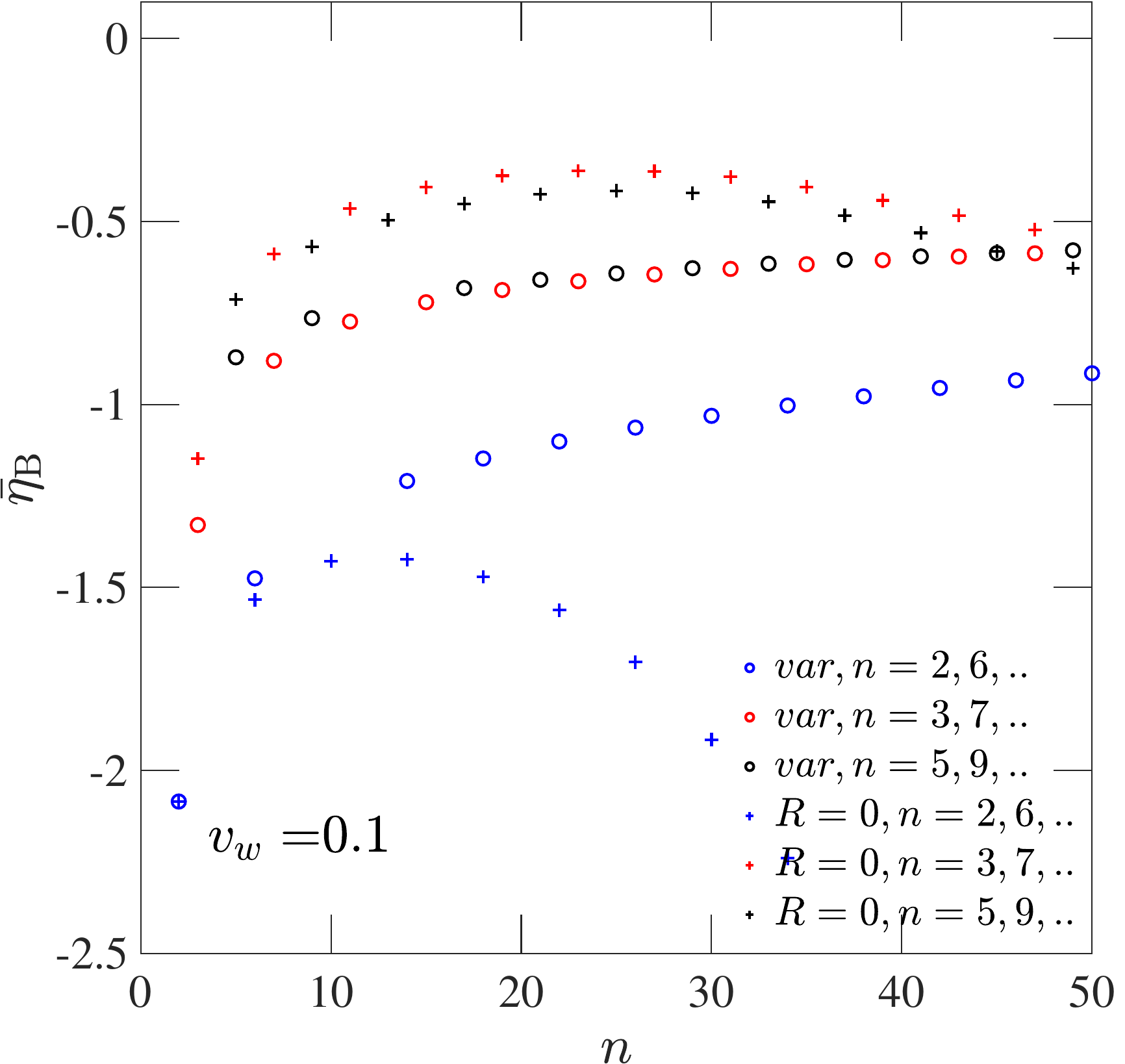}\hskip 0.5truecm
\caption{We show the predicted baryon asymmetry as a function of $n$ for $v_w=0.5$ (upper panels) and for $v_w=0.1$ (lower panels), for variance truncation scheme (open circles) and constant truncation scheme with $R=0$ (plus signs). Left panels show all sequences including $n_i=4$ and the right panels with $n_i=4$ sequence excluded.}
\label{fig:different_sequencies}
\end{figure}

%==================================================================================================
%==================================================================================================

%%%%%%%%%%%%%%%%%%%%%%%%%%%%%%%%%%%%%%%%%%%%%%%%%%%%%%%%%%%%%%%%%%%%%%%%%%%%%%%%%%%%%%%%%%%%%%%%%%%
%
\subsection{Other moment sequences}
\label{sec:moment_sequences}
%
%%%%%%%%%%%%%%%%%%%%%%%%%%%%%%%%%%%%%%%%%%%%%%%%%%%%%%%%%%%%%%%%%%%%%%%%%%%%%%%%%%%%%%%%%%%%%%%%%%%

Let us finally consider moments not belonging to sequence $n=2,6,10,...$ that was used in all above examples. It turns out that there is additional scatter in the final baryon asymmetry, that strongly depends on the sequence used. The three sequences $n = n_i + 4k$, with $k=0,1,...$ and $n_i = 2,3,5$ each separately predict a smooth baryon asymmetry as a function of $k$, but the sequence starting with $n_i=4$ is anomalous, and shows a more erratic behaviour. 

We show our results in figure~\cref{fig:different_sequencies} for $v_w = 0.5$ (upper panels) and for $v_w=0.1$ (lower panels). Open circles correspond to the variance truncation scheme and plus-signs to the constant $R=0$ scheme. Left panels show the results for all $n$ including the $n_i=4$-sequences, whose erratic behaviour is clearly visible. The cause for this behaviour is again revealed by the eigenvalue analysis. In the middle panel of figure~\cref{fig:eigen_values} we showed the eigenvalues of the asymptotic operator $\cal{X}$ for the case $n=8$, which belongs to the $n_i=4$-sequence. Strikingly, we see several eigenvalues with a large complex part and an essentially vanishing real part. The corresponding eigenmodes are thus rapidly oscillating with an extremely slowly converging amplitude as $|z|\rightarrow \infty$. This pattern repeats for all moments in the $n_i=4$-sequence. We believe that occurrence of these $\Re(\lambda)\approx 0$ solutions is a numerical coincidence, possibly related to the fact that our equation network has four particle species. At any rate, the slow decay of the associated eigensolutions makes equations unstable and one should ignore these moments in the further analysis.

Right panels in figure~\cref{fig:different_sequencies} show our results excluding $n_i=4$-sequences. The remaining sequences indeed each yield a smooth BAU as a function of $k$. In the case $v_w=0.5$ the sequences converge to a common value with a scatter that is not larger than the scatter due to different truncation schemes. In the case $v_w=0.1$ the variance truncation solutions for $n_i=3$ and 5 are almost identical and appear to converge with the $n_i=2$-sequence, albeit with $n$ much larger than 50. Moreover, the constant truncation scheme sequences with $n_i=3,5$ are  now in reasonable agreement with the variance truncation schemes, while they both differ from the constant truncation $n_i=2$-sequence, showing that the latter results are again apparently anomalous.

These results confirm that the main source of uncertainty in the moment expansion method arises from the truncation dependence. This is apparently associated with the structure of the asymptotic differential operator at large distances $|z|\rightarrow \infty$, and the constant truncation schemes appear to be more fragile than is the variance truncation scheme. This is presumably due to the fact that the variance scheme relates the moment $u_{n}$ to all lower moments instead of the last included moment $u_{n-1}$, as do the constant truncation schemes. Anyway, the variance truncation scheme appears to work rather well in all cases we considered.

%%%%%%%%%%%%%%%%%%%%%%%%%%%%%%%%%%%%%%%%%%%%%%%%%%%%%%%%%%%%%%%%%%%%%%%%%%%%%%%%%%%%%%%%%%%%%%%%%%%
\paragraph{Discussion.}
%%%%%%%%%%%%%%%%%%%%%%%%%%%%%%%%%%%%%%%%%%%%%%%%%%%%%%%%%%%%%%%%%%%%%%%%%%%%%%%%%%%%%%%%%%%%%%%%%%%

Some robust results emerge from our analysis: firstly, the higher moment calculations tend to give smaller final baryon asymmetry than does the simple 2-moment approximation, in particular for large wall velocities. This is of course not encouraging for the model building efforts trying to realize a sufficient BAU together with a strong electroweak phase transition capable of creating a strong gravitational wave signal. Second, for $v_w \gsim 0.4$, the scatter in the results with different $n$ and different truncation methods is relatively small, of order $\Delta \bar \eta_{\rm B} \approx 0.1$, which is small in comparison to the difference between the 2-moment result and the asymptotic value. 
Finally, for small $v_w$ the overall dependence on $n$ is smaller, and the variance scheme appears to give quite reliable results also here. 

To conclude, let us point out that above we concentrated only on the uncertainty due to moment expansion solution to {\em a given approximation} to the full semiclassical Boltzmann equation. The approximations used to get that equation, and in particular the rather crude estimates for the scattering rates, are potentially a larger source of uncertainty for BAU than are the effects discussed above. For example, simply restoring the nontrivial $\kappa_a$-factors to elastic rates (remember that we used $\kappa_a = 1$ here) causes a change in BAU that is as large as the scatter caused by the different truncation choices. We shall study these issues more quantitatively in~\cite{KimmoNiyati2}.

%%%%%%%%%%%%%%%%%%%%%%%%%%%%%%%%%%%%%%%%%%%%%%%%%%%%%%%%%%%%%%%%%%%%%%%%%%%%%%%%%%%%%%%%%%%%%%%%%%%
%%%%%%%%%%%%%%%%%%%%%%%%%%%%%%%%%%%%%%%%%%%%%%%%%%%%%%%%%%%%%%%%%%%%%%%%%%%%%%%%%%%%%%%%%%%%%%%%%%%
%
\section{Conclusions}
\label{sec:conclusions}
%
%%%%%%%%%%%%%%%%%%%%%%%%%%%%%%%%%%%%%%%%%%%%%%%%%%%%%%%%%%%%%%%%%%%%%%%%%%%%%%%%%%%%%%%%%%%%%%%%%%%
%%%%%%%%%%%%%%%%%%%%%%%%%%%%%%%%%%%%%%%%%%%%%%%%%%%%%%%%%%%%%%%%%%%%%%%%%%%%%%%%%%%%%%%%%%%%%%%%%%%

This article extends the commonly used moment expansion method for solving the semiclassical Boltzmann equations (SCBE's) for the electroweak baryogenesis problem to an arbitrary number of moments $n$. The extension is a fairly straightforward generalization of the 2-moment results of~\cite{Fromme:2006wx,Cline:2020jre}, with some clarifications and corrections. The main computational task is the preparation of a large number of $z$-, $v_w$- and $n$-dependent interpolating functions appearing as coefficients in the general moment equation~\cref{eq:deqs} network. Obtaining closure in this network requires certain factorization and truncation approximations, and it turns out that the resulting baryon asymmetry of the Universe (BAU) can be rather sensitive to the latter of these. 

We find that moment expansion does not behave very well for some truncation choices and some numbers of moments. Perhaps accidentally, in our test case the moment sequence $n = 4k$, with $k=1,...$ returns erratic values for the BAU, which is associated with the emergence of slowly converging and rapidly oscillating eigenfunctions at large distances from the wall. In addition, the constant truncation schemes used previously in the literature show an additional fragility, becoming strongly sensitive to the remaining moment sequences for small wall velocities. In contrast, the new variance truncation scheme introduced here appears to give a reasonably robust prediction for the BAU in all cases we studied.

For $v_w\gsim 0.4$ we found that the uncertainty in the BAU was, for all truncation methods we studied, of order $\Delta \eta_{\rm B} \sim 0.1\eta_{\rm B,obs}$, while BAU itself tends to be about a factor of 5 smaller than the BAU predicted by the simple two-moment case. This suggests that realising both a sufficient BAU and a strong gravitational wave signal might be more challenging than previously thought. For small $v_w$, only the variance truncation scheme gives robust predictions. Here the convergence of the expansion is slower than with large $v_w$, but the overall dependence on $n$ is smaller on the other hand.  We thus believe that the moment expansion in the variance truncation scheme can give a rather accurate approximation to the underlying SC-Boltzmann equation in most cases of interest.

We also pointed out that the final baryon asymmetry may be sensitive to the hydrodynamics of the phase transition, because the shock reheating may change the terminal wall velocities in an inhomogeneous way~\cite{Hindmarsh:2017gnf,Cutting:2019zws,Cutting:2020nla}. Since the baryon number can depend quite strongly on $v_w$, it would be desirable to compute the total baryon number as a weighted integral $\eta_B = \int_0^1 {\rm d}v_w P(v_w) \eta_B(v_w)$, where $P(v_w){\rm d}v_w$ is the fraction of the total volume swept by a transition wall moving with velocity $v_w$. 

%%%%%%%%%%%%%%%%%%%%%%%%%%%%%%%%%%%%%%%%%%%%%%%%%%%%%%%%%%%%%%%%%%%%%%%%%%%%%%%%%%%%%%%%%%%%%%%%%%%%
%%%%%%%%%%%%%%%%%%%%%%%%%%%%%%%%%%%%%%%%%%%%%%%%%%%%%%%%%%%%%%%%%%%%%%%%%%%%%%%%%%%%%%%%%%%%%%%%%%%%
%
\section*{Acknowledgements}
\label{sec:ack}
%
%%%%%%%%%%%%%%%%%%%%%%%%%%%%%%%%%%%%%%%%%%%%%%%%%%%%%%%%%%%%%%%%%%%%%%%%%%%%%%%%%%%%%%%%%%%%%%%%%%%%
%%%%%%%%%%%%%%%%%%%%%%%%%%%%%%%%%%%%%%%%%%%%%%%%%%%%%%%%%%%%%%%%%%%%%%%%%%%%%%%%%%%%%%%%%%%%%%%%%%%%

We thank Jim Cline for discussions and for useful comments on the manuscript. This work was supported by the Academy of Finland grant 318319. NV was in addition supported by grants from the Magnus Ehrnrooth Foundation, the Ellen and Artturi Nyyssönen Foundation, the Vilho, Yrjö and Kalle Väisälä Foundation and the University of Jyv\"askyl\"a Postgraduate School.

%%%%%%%%%%%%%%%%%%%%%%%%%%%%%%%%%%%%%%%%%%%%%%%%%%%%%%%%%%%%%%%%%%%%%%%%%%%%%%%%%%%%%%%%%%%%%%%%%%%
%%%%%%%%%%%%%%%%%%%%%%%%%%%%%%%%%%%%%%%%%%%%%%%%%%%%%%%%%%%%%%%%%%%%%%%%%%%%%%%%%%%%%%%%%%%%%%%%%%%
%
\section*{Appendices}
\addcontentsline{toc}{section}{\protect\numberline{}Appendices}
\appendix
%
%%%%%%%%%%%%%%%%%%%%%%%%%%%%%%%%%%%%%%%%%%%%%%%%%%%%%%%%%%%%%%%%%%%%%%%%%%%%%%%%%%%%%%%%%%%%%%%%%%%
%%%%%%%%%%%%%%%%%%%%%%%%%%%%%%%%%%%%%%%%%%%%%%%%%%%%%%%%%%%%%%%%%%%%%%%%%%%%%%%%%%%%%%%%%%%%%%%%%%%
%
\section{Explicit moment functions}
\label{sec:app_explicit_forms}
%
%%%%%%%%%%%%%%%%%%%%%%%%%%%%%%%%%%%%%%%%%%%%%%%%%%%%%%%%%%%%%%%%%%%%%%%%%%%%%%%%%%%%%%%%%%%%%%%%%%%
%%%%%%%%%%%%%%%%%%%%%%%%%%%%%%%%%%%%%%%%%%%%%%%%%%%%%%%%%%%%%%%%%%%%%%%%%%%%%%%%%%%%%%%%%%%%%%%%%%%

As explained in~\cite{Cline:2020jre}, all coefficient functions appearing in the moment equations~\cref{eq:genericmomeq} can be written in terms of a single dimensionless integral function
\begin{equation}
K({\cal F}_0;V;n,m)
 = -\frac{3}{\pi^2\gamma_w}\int_x^\infty {\rm d}w \int_{-1}^1 {\rm d}y \times \,\frac{\tilde p_w\tilde p_z^n}{\tilde E^{m-1}}\, V(w,y,v_w,x)\,{\cal F}_0(w),
\label{eq:Kfun}
\end{equation}
where ${\cal F}_{0w} = f_{0w}, f^\prime_{0w}$ or $f^{\prime\prime}_{0w}$ and $\tilde p_w \equiv \sqrt{w^2\!-\!x^2}$, $\tilde p_z = \gamma_w(y\tilde p_w-wv_w)$ and $\tilde E = \gamma_w(w-v_wy\tilde p_w)$. For $D_\ell$, $Q_\ell$, $Q^e_\ell$ and for $K_1$ the auxiliary function $V \!=\! 1$. For the source functions $Q^{8o}_\ell$ and $Q^{9o}_\ell$ a nontrivial structure $V = s_{\rm p}p_z/{E_z}$ appears. For the spin $s$ eigenstates then:
\begin{equation}
  V = V_s = \frac{|p_z|}{E_z} = \frac{|\tilde p_z|}{\sqrt{\tilde p_z^2+x^2}}
\end{equation}
while for helicity eigenstates:
\begin{equation}
  V = V_h \equiv V_s^2 \left(1 - \frac{m^2}{\tilde E^2} \right)^{\!-1/2}.
\end{equation}
where $\tilde p_z$ and $\tilde E$ are as given below~\eqref{eq:Kfun}. Explicitly then:
\begin{align}
  D_\ell(x,v_w) &=  K(f'_0;1;\ell,\ell ) \nn\\
  Q_\ell(x,v_w) &=  K(f''_0;1;\ell-1,\ell ) \nn\\
  Q^{8o}_\ell(x,v_w) &= \sfrac{1}{2} K(f'_0;V;\ell-2,\ell ) \nn\\
  Q^{9o}_\ell(x,v_w) &= \sfrac{1}{4} \big[ K(f'_0;V;\ell-2,\ell+2)
                                  -\gamma_w K(f_0'';V;\ell-2,\ell+1)\big].
\label{eq:moment_functions_appendix}                                  
\end{align}
The function $\bar R$ is a special case, whose one-dimensional integral representation was already given in~\eqref{eq:barR}.

%%%%%%%%%%%%%%%%%%%%%%%%%%%%%%%%%%%%%%%%%%%%%%%%%%%%%%%%%%%%%%%%%%%%%%%%%%%%%%%%%%%%%%%%%%%%%%%%%%%
%%%%%%%%%%%%%%%%%%%%%%%%%%%%%%%%%%%%%%%%%%%%%%%%%%%%%%%%%%%%%%%%%%%%%%%%%%%%%%%%%%%%%%%%%%%%%%%%%%%
%
\bibliography{CKV.bib}
%
%%%%%%%%%%%%%%%%%%%%%%%%%%%%%%%%%%%%%%%%%%%%%%%%%%%%%%%%%%%%%%%%%%%%%%%%%%%%%%%%%%%%%%%%%%%%%%%%%%%
%%%%%%%%%%%%%%%%%%%%%%%%%%%%%%%%%%%%%%%%%%%%%%%%%%%%%%%%%%%%%%%%%%%%%%%%%%%%%%%%%%%%%%%%%%%%%%%%%%%

%%%%%%%%%%%%%%%%%%%%%%%%%%%%%%%%%%%%%%%%%%%%%%%%%%%%%%%%%%%%%%%%%%%%%%%%%%%%%%%%%%%%%%%%%%%%%%%%%%%
%%%%%%%%%%%%%%%%%%%%%%%%%%%%%%%%%%%%%%%%%%%%%%%%%%%%%%%%%%%%%%%%%%%%%%%%%%%%%%%%%%%%%%%%%%%%%%%%%%%
%
\end{document}